\documentclass[twocolumn,pre,superscriptaddress,longbibliography]{revtex4-2}

\usepackage{graphicx}
\usepackage{amsmath}
\usepackage{dcolumn}
\usepackage{bm}
\usepackage{ulem}
\usepackage{xcolor}

\begin{document}

\newcommand{\be}{\begin{equation}}
\newcommand{\ee}{\end{equation}}
\newcommand{\bea}{\begin{eqnarray}}
\newcommand{\eea}{\end{eqnarray}}
\newcommand{\bse}{\begin{subequations}}
\newcommand{\ese}{\end{subequations}}
\newcommand{\comment}[1]{}
\newcommand{\av}[1]{\left\langle #1 \right\rangle}
\newcommand{\e}{\varepsilon}              %

\newcommand{\AP}[1]{{\color{blue} [AP: #1]}}
\newcommand{\PP}[1]{{\color{red}[PP: #1]}}
\newcommand{\SL}[1]{{\color{teal}[SL: #1]}}

\newcommand{\new}[1]{{\color{magenta} #1}}
\renewcommand{\e}{\epsilon}

\title{Lattice models of random advection and diffusion and their statistics}
\author{Stefano Lepri}
\affiliation{Consiglio Nazionale delle Ricerche, Istituto dei Sistemi Complessi, Via Madonna del Piano 10 I-50019 Sesto Fiorentino, Italy.} 
\affiliation{Istituto Nazionale di Fisica Nucleare, Sezione di Firenze, 
via G. Sansone 1, I-50019 Sesto Fiorentino, Italy}
\author{Paolo Politi}
\affiliation{Consiglio Nazionale delle Ricerche, Istituto dei Sistemi Complessi, Via Madonna del Piano 10 I-50019 Sesto Fiorentino, Italy.} 
\affiliation{Istituto Nazionale di Fisica Nucleare, Sezione di Firenze, 
via G. Sansone 1, I-50019 Sesto Fiorentino, Italy}
\author{Arkady Pikovsky}
\affiliation{Department of Physics and Astronomy, University of Potsdam
Karl-Liebknecht-Str 24/25, Bldg 28    D-14476, Potsdam, Germany}

\begin{abstract}
We study in detail a one-dimensional lattice model of a continuum, conserved field (mass)
that is transferred deterministically between neighbouring random sites.
The model falls in a wider class of lattice models capturing the joint effect of
random advection and diffusion and encompassing as
specific cases, some models studied in the literature,  
like the Kang-Redner, Kipnis-Marchioro-Presutti, Takayasu-Taguchi, etc. 
The motivation for our setup comes from a straightforward interpretation as
advection of particles in one-dimensional turbulence, but it is also related
to a problem of synchronization of dynamical systems driven by common noise. 
For finite lattices, we study both the coalescence of an initially spread field
(interpreted as roughening), and the 
statistical steady-state properties.  
We distinguish two main size-dependent regimes, depending on the strength of the diffusion term 
and on the lattice size.  Using numerical simulations and mean-field
approach, we study the statistics of the field.
For weak diffusion, we unveil a characteristic
hierarchical structure of the field. We also connect the model and the iterated function systems concept.
\end{abstract}

\maketitle 
\section{Introduction}
\label{sec:intro}

Advection and diffusion are two basic transport phenomena occurring
in diverse physical contexts. The former amounts to the motion of, for
instance, small tracer particles (like a pollutant) transported by
the movement of a surrounding fluid. On the other hand, diffusion is caused by the familiar
mechanism of Brownian random walk that causes a stochastic spreading of
tracer particles due to the interaction with a solvent.

A particularly interesting case is where advection
is a random process. In the applications 
to turbulence, there is a vast literature on the matter \cite{warhaft2000passive}. 
A celebrated example is the Kraichnan model for the advection of a passive 
scalar by a random flow \cite{kraichnan1994anomalous}.
In this case,  one usually assumes an incompressible 
(solenoidal) velocity field
\cite{kraichnan1994anomalous} or a weakly compressible 
fluid~\cite{Elperin_etal-95}.  As it is known, the former case is 
related to Hamiltonian dynamical systems theory,  as exemplified
by Lagrangian chaos \cite{vulpiani2009chaos}. On the other hand, the issue of compressible fluids is 
less studied and corresponds to dissipative phase-space flows.

Besides the problem of passive scalar transport in fluids,
the concept of advection is more general and applies in more
abstract sense to spreading of an ensemble of trajectories in phase space
of a dynamical system subject to a common regular or irregular phase velocity field.
Examples of this setup appear in neurosciences and 
other fields.
Another interesting application concerns transport in active 
media  \cite{deutsch1994probability} as it occurs for
light in disordered and amplifying systems \cite{deutsch1994probability,lepri2013fluctuations}.

Mathematically, a description of the problem in the continuum limit
requires dealing with a stochastic partial differential equation, that
are notoriously hard to deal with. From a more statistical-mechanics
point of view, it is thus helpful to consider simple \textit{discrete} 
microscopic or
mesoscopic models of the dynamics that respect  some fundamental
features of the problem. Such an approach is insightful as it allows
to simulate the process straightforwardly.
In this work, we follow this strategy to take a fresh look at 
the problem where random advection and diffusion are \textit{both} present. 
We introduce a general
class of stochastic lattice models where microscopic moves  mimic
the two basic mechanisms, namely the collective random motion of particles
induced by the common advecting field and the spreading caused by
microscopic diffusion (this distinction will be made clear in the
following). For simplicity, we deal with a one-dimensional lattice.
Discrete dynamics is easily generalized to higher dimensions or graphs, although
in these cases a relation to the original continuous advection setup becomes non-trivial (see discussion in Scetion~\ref{sec:cscm}).
We anticipate  the class  to encompass various models considered
previously in the literature as particular cases.

The primary model we are going to study depends on a single parameter
$\e$ confined to the unitary interval $[0,1]$ and
allowing for tuning the relative importance of advection and diffusion.
This parameter quantifies the fraction of mass which is
transferred from a random site to a random neighbour.
In one limit ($\epsilon =0$) there is a whole transfer of mass and
the process is characterized by macrodiffusion (or random advection).
In the opposite limit ($\epsilon\to 1$) a vanishing piece of matter
is transferred and the process is characterized by microdiffusion.
For general $\epsilon$, both processes are present.

We are interested both in the time dependence of the field 
evolving from an initial
uniform state and in the properties
of the statistically stationary state that emerges at large times. 
In the former case, the typical phenomena are coarsening and roughening, 
namely how clusters merge and how the field variance grows in time. 
Also, the steady-state statistics of the field is of great interest.
We will report cases where the statistics is strongly non-Gaussian
and we will be able to give a scaling description of the asymptotic state
for any $\e$ and any size $L$ of the system. 

The paper is organized as follows. In Section \ref{sec:general} we describe
the basic phenomenology of advection and diffusion in a smooth, random
field and introduce the basic distinction between macroscopic 
(collective) and microscopic diffusion.  
The general class of models with stochastic microscopic 
dynamics is defined in Section \ref{sec:lattice}. Particular
cases corresponding to various systems studied in the 
literature are examined there. 
Our analysis starts by considering the case of no microscopic 
diffusion for a finite lattice (Section \ref{sec:kr}) 
and its roughening properties.  We then 
focus on various steady-state properties in 
Sections \ref{sec:ttstat},\ref{sec:lae} and \ref{sec:sme}.
Conclusions are given in Section \ref{sec:cscm}, along with 
a brief comparison of our results with those
given in the literature for other models with similar
conservation laws.  
Some more technical aspects are relegated to the Appendix.

\section{Phenomenology of random one-dimensional advection and diffusion}
\label{sec:general}

Let us start by discussing general qualitative concepts about one-dimensional random passive scalar advection. The starting point is an ensemble of ``particles'' with coordinates $x_i(t)$ in a velocity field $v(x,t)$, so that the dynamics is simple
\begin{equation}
\frac{d x_i}{dt}=v(x_i,t)\;.
\label{eq:rae}
\end{equation}
We assume that the velocity field $v(x,t)$ is a random function of time so that each particle displays a random one-dimensional motion. In order for the differential equation \eqref{eq:rae} to be well-posed, the field $v(x,t)$ should be smooth enough in $x$ (at least Lipschitz continuous). 

One typically associates advection with a transport by moving fluid, and fluids are in most cases nearly incompressible, so that in one dimension the velocity is constant. However, the velocity field on a surface of incompressible fluid can be any function of coordinate and time. Thus, the one-dimensional setup directly applies to particles floating on the surface of a two-dimensional turbulent flow. This flow can be random surface waves (see \cite{PhysRevLett.76.4717}  for experiments involving particles' advection in two-dimensional surface waves and~\cite{Ricard_2021} for 
recent realization of turbulent one-dimensional surface waves). Another possible source
of one-dimensional random advection is two-dimensional turbulent convection (e.g., in a Hele-Show cell) with an
open upper surface, on which the floating particles move.

The main macroscopic effect is the merging/clustering/coalescence (we use these terms as synonyms below) of particles. Indeed, if the coordinates of two particles coincide at $t=0$: $x_i(0)=x_j(0)$, then their trajectories are identical at any later time,  i.e., $x_i(t)\equiv x_j(t) =X(t)$ $\forall t>0$,  with $\dot X=v(X,t)$ and $X(0)=x_{i,j}(0)$. A cluster of particles with identical positions is, therefore, a solution of Eq.~\eqref{eq:rae}. A second and equally important remark is that such a solution is stable in the sense that neighboring particles get effectively ``attracted'' to each other to form a cluster. To see this, suppose that the field $v(x,t)$ is a smooth enough  function of $x$, so that one can linearize \eqref{eq:rae} around a reference trajectory of a cluster $X(t)$, to obtain for a small perturbation $\Delta x$
\begin{equation}
\frac{d}{dt} \Delta x=\Delta x \frac{\partial}{\partial x} v(X(t),t)\;.
\label{eq:linrae}
\end{equation}
This linear equation with a random function of time $v(X(t),t)$ defines the Lyapunov exponent
\begin{equation}
\lambda=\av{\frac{\partial}{\partial x} v(X(t),t)}_t
\label{eq:le}
\end{equation}
so that asymptotically in time, $\Delta x (t)\propto \exp[\lambda t]$. 

The main observation is that in one-dimensional continuous dynamics, the Lyapunov exponent cannot be positive,
because the phase volume for a statistically stationary regime cannot grow indefinitely.
Furthermore, it is unprobable for random fields that the Lyapunov exponent vanishes. 
Indeed, in  non-random one-dimensional dynamics, the Lyapunov exponent can be either negative (a sink)
or zero (e.g., a steady periodic motion over a periodic space profile). Randomness
``mixes'' these two situations, thus leading to a negative Lyapunov exponent.
For a negative Lyapunov exponent, $\Delta x\to 0$ as $t\to\infty$. 
This means that neighboring particles glue together (coalesce)
and in a finite system, a stable cluster forms at long enough times.
Since there \textit{all} the particles have the same 
trajectory, we will refer to it as the \textit{maximal cluster}.
Also, since the cluster will perform a random motion, we will, for definiteness, refer to 
this motion as \textit{macrodiffusion}. 

We note here that coalescence to a maximal cluster also occurs in more general situations,
provided the maximal Lyapunov exponent is negative. For example, such a situation
is possible for two-dimensional advection as well, although in this case, there are
two Lyapunov exponents, so the maximal one may become positive, and the cluster will be destroyed.
One class of problems where a maximal cluster appears is irreversible aggregation,
where a large number of small particles coalesce over time with no 
possibility of breaking up, see Ref.~\cite{leyvraz2003scaling} 
and the literature therein. 
On the other hand, in the
context of noise-driven dynamical systems,
the effect of the formation of the maximal cluster 
has been termed \textit{synchronization by common noise} and  
was first described in Refs.~\cite{Pikovsky-84a,antonov1984}, 
where an ensemble of identical systems (i.e., an ensemble of different initial conditions) driven by
the same realization of noise was analyzed. In mathematical literature, equations
of type  \eqref{eq:rae} are called random dynamical systems and a maximal cluster state as described above 
represents a point random attractor in such a system~\cite{Crauel-Flandoli-94}. The effect of synchronization
by common noise also appears in neuroscience (there it is called reliability \cite{Mainen-Sejnowski-95}),
and in other fields \cite{Khoury-98,Uchida-Mcallister-Roy-04}.
If the maximal Lyapunov exponent becomes positive (what is possible starting from dimension two), 
a point attractor undergoes a transition to a fractal one~\cite{Pikovsky-92c,Sommerer-Ott-93a}. In the context
of passive scalar advection theory, such a transition, happening as compressibility of the underlying flow increases,
was discussed in \cite{Gawedzki-Vergassola-00}.

The nature of the macrodiffusion (i.e.,  whether it is
normal or anomalous) depends on the actual statistical properties of the field $v(x,t)$ 
(cf.~\cite{Bohr-Pikovsky-93}). 
In the examples considered below, we will limit to the case where
spatial and temporal correlations of the velocity field decay rapidly, 
so that the macrodiffusion will be normal.

We stress that the arguments about attracting clustered states are valid for a large but finite system.
This will be the case we consider in this paper, where we also will explore scaling relations as the size
of the system goes to infinity.

Although a clustered state is an attractor in the random advection dynamics \eqref{eq:rae},
there can be several such attractors in degenerate situations. Indeed, if, e.g., the velocity field is odd in space $v(-x,t)=-v(x,t)$,
then $v(0,t)=0$ and the particles starting in positive $x>0$ and negative $x<0$ domains never meet and never merge,
so there are at least two attractors here. Below, we assume that such a degenerate situation does not occur, and we 
have one ergodic component: all the initially distributed particles evolve under \eqref{eq:rae} 
to one single maximal cluster.

The above concept of macrodiffusion should be juxtaposed with a familiar Brownian motion, 
as given by the Langevin equations
\begin{equation}
\frac{d x_i}{dt}=\sigma\xi_i(t)\;,
\label{eq:brown}
\end{equation}
where the microscopic noises $\xi_i(t)$ are zero-average, independent (i.e., different for different particles) 
Gaussian and white noises with 
$\av{\xi(t)}=0$, $\av{\xi_i(t)\xi_j(t')}=2\delta_{ij}\delta(t-t')$.
This noise leads to a diffusive spreading $\sim \sigma^2 t$ of particles in the ensemble.
We call this effect \textit{microdiffusion} for a clear distinction with the above.

\begin{figure}[!htb]
\centering
\includegraphics[width=\columnwidth]{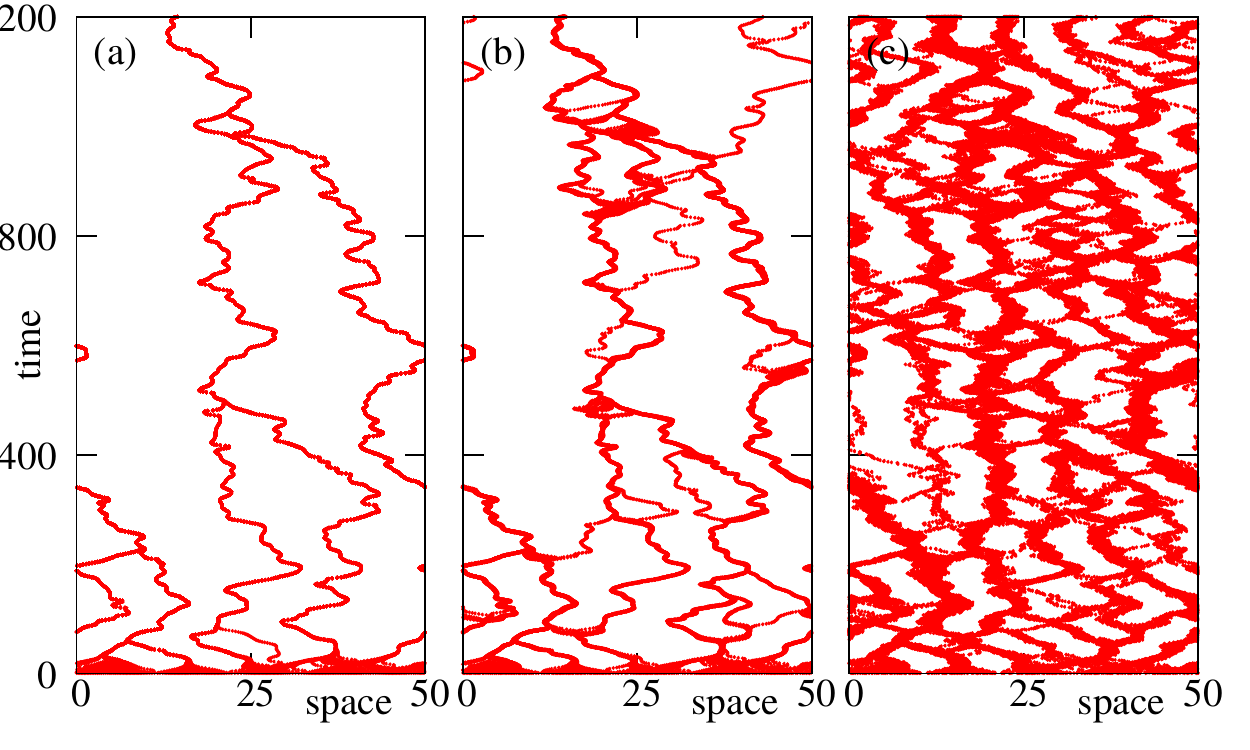}
\caption{Dynamics of a set of 200 particles, initially uniformly distributed, according
to Eq.~\eqref{eq:rane} for different levels of microdiffusion $\sigma$, for the same realization
of random field $v(x,t)$ (this field is taken as a chaotic solution of the Kuramoto-Sivashinsky 
equation, cf.~\cite{Bohr-Pikovsky-93}). Panel (a): microdiffusion-free case $\sigma=0$, here one observes a perfect formation
of the maximal cluster. Panel (b): small noise $\sigma=0.005$, here some particles due to noise split from
the largest cluster, which, however, contains most of them. Panel (c): large noise $\sigma=0.3$, here the distribution
of particles is non-uniform, but there are no dominating clusters.}
\label{fig:rane}
\end{figure}

Our main goal in this paper is to contribute to an understanding of the behavior of an ensemble of particles, where both macro- and microdiffusion  in a given random velocity field are present; namely, we combine \eqref{eq:rae} and \eqref{eq:brown} into an equation
\begin{equation}
\frac{d x_i}{dt}=v(x_i,t)+\sigma\xi_i(t).
\label{eq:rane}
\end{equation}
We illustrate different regimes in the dynamics of \eqref{eq:rane} in Fig.~\ref{fig:rane}.

Equation \eqref{eq:rane} can be considered as a Langevin equation so that the evolution of 
the probability density $w(x,t)$ of an ensemble of particles obeys the Fokker-Planck equation
\begin{equation}
\frac{\partial}{\partial t}w+\frac{\partial}{\partial x} \left(v(x,t)w\right)=
\sigma^2 \frac{\partial^2 w}{\partial x^2}\; .
\label{eq:ade}
\end{equation}
We stress that this equation includes averaging over the microscopic noise terms $\xi(t)$ but still contains
the random function $v(x,t)$, which is common for all the particles. Thus, formally this equation
is a stochastic partial differential equation.

Our main interest is in the statistical properties of the density $w(x,t)$.
As argued above, in the absence of microdiffusion, $\sigma=0$, 
the asymptotic state is a point attractor (cluster)
which moves randomly: $w(x,t)=\delta(x-X(t))$. This singular 
solution becomes ``smeared'' by a finite microdiffusion $\sigma>0$. 
However, the details of this continuous random field $w(x,t)$
are not clear a priori, and the goal of this paper is to contribute to understanding 
the statistical properties of the distribution density $w$.

\section{Lattice models}
\label{sec:lattice}

It is computationally rather costly to simulate Eq.~\eqref{eq:ade} 
on a large domain and for small $\sigma$. Furthermore, the results are expected to depend drastically on statistical characteristics of the underlying velocity field $v(x,t)$. In the literature, one has taken for this field turbulent solutions of the deterministic
Kuramoto-Sivashinsky equation~\cite{Bohr-Pikovsky-93,PhysRevE.49.5853} as in Fig.~\ref{fig:rane}. Another
approach is to interpret the advective motion as sliding along a one-dimensional surface $v(x,t)=\partial_x h(x,t)$, and to use for the dynamics of this surface $h(x,t)$ one of the popular stochastic partial differential  equations, e.g., the Edwards-Wilkinson equation~\cite{Edwards-Wilkinson-82} or the Kardar-Parizi-Zhang equation~\cite{PhysRevLett.56.889}, see refs.~\cite{PhysRevLett.85.1602,PhysRevB.66.195414,PhysRevE.64.046126,PhysRevE.66.021104,PhysRevE.74.021124,PhysRevE.98.052148}. Although there are statistical models for one-dimensional wave turbulence~\cite{zakharov2004one}, including the deterministic fractional PDE by Majda, McLaughlin, and Tabak~\cite{MMT}, we are not aware of any study of advection in such models.

A convenient approach to finding scaling properties of random advection is to 
explore proper lattice models. 
In a lattice model, the field is discrete in space, and therefore one cannot perform a continuous stability
analysis of a cluster state like in Eq.~\eqref{eq:linrae} above. Indeed, if one takes the continuous in space
model~\eqref{eq:rae}, then a natural assumption is that the velocity fields at large distances are independent, but
at small distances (at which the linearization \eqref{eq:linrae} is valid) it is smooth. Thus, if one takes
two particles at a large initial distance, they first diffuse independently, 
and only when they are close enough to each other
do they merge according to the Lyapunov exponent \eqref{eq:le}. 
A lattice model can imitate the first stage of independent diffusion but 
replaces the second stage of exponential convergence with an abrupt coalescence. Furthermore,
the lattice models below are formulated in a discrete time. In these models one defines 
a conserved ``mass field'' $u_k(t)$, where $k$
is the lattice site, and $t$ is discrete time. This field should be interpreted as a 
discretized density $w(x,t)$ of 
advected particles from
equation \eqref{eq:ade}. As it is clear from the discussion of existing continuous models above, one can construct lattice models with different statistical characteristics. Below we will focus on ``maximally random'' lattice models, with vanishing correlations of the effective ``velocity field'' in space and time. The motion of a single
particle in such models is pure diffusion, in contradistinction to a superdiffusion due to time-correlation of the velocity in the cases mentioned above~\cite{Bohr-Pikovsky-93,PhysRevE.49.5853,PhysRevE.66.021104}.

\subsection{Generic two-site models}
\label{sec:dm}

We start with a rather generic setup and then focus on two particular lattice models that will be considered below.
We assume that the models for a continuous field on a lattice are formulated as follows:
\begin{enumerate}
\item A pair of neighboring (to ensure locality) sites, $(k,k+1)$, is chosen randomly.
\item The fields at these sites, $u_k,u_{k+1}$, are redistributed according to a deterministic (parameters fixed) or stochastic 
(parameters chosen from a distribution) linear rule,
\bea
u'_k &=& (1-a) u_k + b u_{k+1} \\
u'_{k+1} &=& a u_k + (1-b) u_{k+1},
\eea
where the primes indicate the masses after a move and where we have implemented mass conservation.
\end{enumerate}
Therefore, the rule is generically described by a stochastic matrix depending
on two parameters $0\leq a,b\leq 1$:
\[
\mathsf{A}(a,b)=\begin{pmatrix} 1-a& b\\ a &1-b\end{pmatrix}\;.
\]
If $a\neq b$, the distribution is asymmetric. Thus, for fixed $a\neq b$, one applies matrices $\mathsf{A}(a,b)$ and
$\mathsf{A}(b,a)$ with probabilities $1/2$. Such a symmetrization might not be needed if $a,b$ are chosen as random variables. We note that the parameters  $k,k+1$ and matrix  $\mathsf{A}(a,b)$ do not depend on the
field $u$, thus the particles are \textit{passive}. For a lattice model of an \textit{active} particle sliding over a surface which is influenced by the particle position, see~\cite{PhysRevE.99.042124}.

 \begin{figure}[!htb]
 \centering
 \includegraphics[width=0.65\columnwidth]{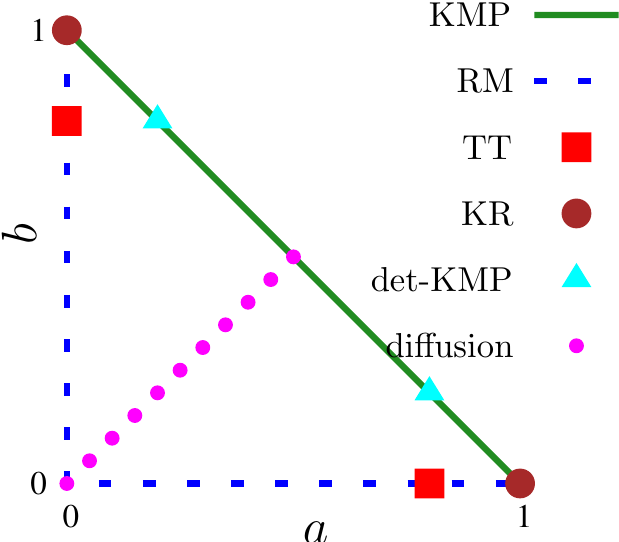}
 \caption{Different models in terms of parameters $(a,b)$. Cases TT, KR, and det-KMP with fixed parameters  (except for diffusion) are shown with markers (for definiteness, we take $\e=0.8$);
 cases KMP and RM, where parameters are random, are shown with green solid and blue dashed lines, respectively, 
 on which the values of these parameters lie. The case of diffusion is shown with a magenta dotted line.
 }
 \label{fig:sk} 
 \end{figure}

 In terms of these parameters, different models from the literature can be described as follows:
 \begin{itemize}
 \item \textbf{Kang-Redner (KR) model}~\cite{kang1984fluctuation} (see a detailed description in 
 section \ref{sec:krm} below) corresponds to $a=1,b=0$:
all the mass of a random site $i$ is transferred to a random neighbour $i\pm 1$.
 \item \textbf{Takayasu-Taguchi (TT) model}~\cite{takayasu1993non}  (see a detailed description in  section \ref{sec:tt} below) corresponds to $a=1-\e$, $b=0$:
the fraction $a$ of the mass of a random site $i$ is transferred to a random neighbor $i\pm 1$.
 \item \textbf{Kipnis-Marchioro-Presutti (KMP) model}~\cite{kipnis1982heat} (see discussion in section \ref{sec:cscm}) corresponds to $a=1-b=\xi$, where $\xi$ is uniformly distributed $0\leq \xi\leq 1$:
the total mass of a random pair of neighboring sites is randomly redistributed between them.
 \item \textbf{Rajesh-Majumdar (RM) model}~\cite{Rajesh_Majumdar_2000} (see discussion in section \ref{sec:cscm}) in the limiting case 
 of a sequential update is a random version of TT model $a=\xi$, $b=0$, 
 where $\xi$ is uniformly distributed $0\leq \xi\leq 1$.  
 \item \textbf{Deterministic KMP (det-KMP) model} (it looks like this model has not been considered before)
 corresponds to $a=\e$, $b=1-\e$: a certain portion of the total mass of a random pair of neighboring sites is distributed between them
 in a fixed proportion. 
 \item \textbf{Diffusion} This is a situation when $a=b$ (either fixed or random):
a random site gains (or loses) a fraction of the mass difference between it and a neighboring site. 
 \end{itemize}
In this paper, we focus on the KR and TT lattice models and discuss the relation to KPM, RM, det-KPM, and some models
based on the particle dynamics in section \ref{sec:cscm}.

\subsection{Kang-Redner model}
\label{sec:krm}

This setup is attributed to Smoluchowski; it describes 
coalescence without microdiffusion (i.e., at $\sigma=0$).
We outline it following the paper
\cite{kang1984fluctuation} (where this model is also discussed in dimensions higher than one). 
In this Kang-Redner (KR) model, the field $u_k$ is fully discrete: 
on each site of a discrete, regular one-dimensional
lattice, the mass is an integer number of ``particles'', $u_k=0,1,2,\ldots$. 
Such masses macrodiffuse with a constant diffusion constant, according 
to the following sequential update: at each time step, a site $k$ is chosen at random, 
along with the direction of 
motion ($\pm$, also chosen at random). Then 
the mass migrates from site $k$ to the neighboring site:
\begin{equation}
\begin{aligned}
&u_{k}(t+1)= 0\;,\quad \\
&u_{k\pm 1}(t+1)= u_{k\pm 1}(t)+u_k(t)\;. 
\end{aligned}
\label{eq:kr}
\end{equation}
This model is sometimes called the $A_i+A_j\to A_{i+j}$ kinetic reaction \cite{leyvraz2003scaling}.
Clearly, on a finite lattice of size $L$, the distribution of masses 
converges to a state where
all the particles occupy the same lattice site, and this maximal cluster performs a random walk.

The relaxation dynamics towards such a final state is characterized by the temporal
evolution of the probability $c_m(t)$ to have a cluster of $m>0$ particles.
In one dimension and in the infinite domain, 
the authors of \cite{kang1984fluctuation} provide the 
following scaling relation for $c_m(t)$:
\begin{equation}
\begin{gathered}
c_m(t)\sim  \frac{m}{t^{3/2}}  \, f\left(\frac{m}{t^{1/2}}\right)\;,\quad m>0\;,\\
 \text{where}\quad f(x)\to \begin{cases}
 1 & x\ll 1\;,\\ 0\quad \text{(rapidly)}& x\gg 1\;.\end{cases}
 \end{gathered}
\label{eq:krdist}
 \end{equation}

A similar result, $c_m(t) \sim \frac{m}{t^{3/2}}$ valid for $m\ll t^{1/2}$ is derived in 
\cite{krishnamurthy2003persistence}.
We notice here that one can also formulate the KR model for masses that are not integers, but
any non-negative real numbers (as is, e.g., assumed in the Takayasu-Taguchi model below). 
The phenomenology is the same: in the course of time, masses coalesce, and
in a finite system, eventually, one moving maximal cluster contains all the initial mass. However,
in this case, one has to generalize the discrete distribution $c_m(t)$ into a continuous one; thus,
we stick to the original Kang-Redner formulation for discrete particles.

\subsection{Takayasu-Taguchi model}
\label{sec:tt}

In this work, we will focus on another microscopic model,
first introduced by Takayasu and Taguchi (TT) in Ref. \cite{takayasu1993non}. It  
is defined for a continuous lattice field $u_k(t)$, defined on sites
$k=1,\ldots, L$ (with periodic boundary conditions)
and discrete time, $t=0, 1,2, \ldots$. 
The dynamical rule is very similar to that in the KR model. 
For a randomly chosen site $k$ and a randomly chosen
``direction'' $\pm$, the field is updated as
\begin{equation}
\begin{aligned}
&u_{k}(t+1)= \e u_k(t),\quad \\
&u_{k\pm 1}(t+1)= u_{k\pm 1}(t)+(1-\e) u_k(t)\;. 
\end{aligned}
\label{eq:tt}
\end{equation}
The parameter $0\le \e < 1$ gauges the strength of field transport,
because a fraction $(1-\e)$ of the mass $u_k$ is transported
to a neighboring site and added to $u_{k\pm 1}$~\footnote{%
Our notation is different from the notation of the original paper \cite{takayasu1993non} 
that employed the parameter $j=1-\e$. Our choice is due to our interest towards the limit $\e\to 0$.}
(the case $\e=1$ is trivial because there is no dynamics).
By construction, the dynamics conserves the total mass, $\sum_k u_k$.

The TT dynamics entails both macro- and microdiffusion, 
their relative strengths being controlled by the 
parameter $\e$. This follows from the two interesting limits. 
\begin{enumerate}
\item
For $\e\to 1$
the advection is weak, so the microdiffusion is relatively strong. 
In this case, at each step, a small portion of the field at a randomly chosen site is
transferred left or right~\footnote{At first glance, this statement is not obvious
because the limit $\e\to 1$ might not commute with the thermodynamic limit $L\to\infty$.
However, we will provide a consistent description for any $L$ and $\e$.}.
\item
On the other hand, 
the case $\e=0$ is one without microdiffusion but with pure random advection.
It is also termed irreversible aggregation in Ref.~\cite{takayasu1993non}. 
Here, the fields at neighboring sites 
merge (coalesce), but no further splitting is possible.
In fact, one can easily see that for $\e=0$ the TT model \eqref{eq:tt} 
is essentially the same as the  KR model 
\eqref{eq:kr}, the only difference being the allowed set of values of $u_k$: 
in the KR model these
values are integers, while in the TT model, they are non-negative real numbers.
However, this difference appears to be irrelevant for large systems and for large times when clusters with 
large occupations emerge. 
\end{enumerate}

It is also to be mentioned that some variants of 
the TT model, including external injection, have been considered in Refs.
\cite{takayasu1994non,takayasu1996fractal}.

\subsection{Takayasu-Taguchi model with global coupling}

In the following, we will also consider a variant of 
the TT model with \textit{global} interaction. By this we 
mean that the exchange does not occur between the neighbors 
but rather between two independently randomly chosen sites $k,m$.
Evolution follows the same rule,
\begin{equation}
\begin{aligned}
&u_{k}(t+1)= \e u_k(t)\;,\\
&u_{m}(t+1)= u_{m}(t)+(1-\e) u_k(t)\;. 
\end{aligned}
\label{eq:ttl}
\end{equation}
Remarkably, this model is exactly the multiplicative random exchange model, introduced in \cite{ispolatov1998wealth} to describe wealth redistribution in a population. In general, wealth redistribution models
share some properties, like conservation of the total mass, with the random advection models above, but are typically formulated with discrete agents not on a lattice (so that there is no spatial organization and, correspondingly, no locality) but with global coupling, see~\cite{dragulescu2000statistical,heinsalu2014kinetic,van2016duality}.

\subsection{The TT model as an Iterated Function System}
The TT model has a remarkable mathematical interpretation as
an Iterated Function System (IFS) with probabilities
\cite{barnsley2014fractals}. Indeed, each advection event in \eqref{eq:tt}
is a linear contracting transformation of the vector $\{u_k\}$, and there are altogether
$2L$ different such transformation ($L$ sites multiplied by two possible transport directions).
Thus the probability for one particular transformation is $(2L)^{-1}$. 
Evolution is a composition of these transformations and fulfills the definition
of an Iterated Function System (Chapter IX of book~\cite{barnsley2014fractals}).
Typically, IFSs are used to produce fractal measures. We show in Appendix \ref{sec:bc}
that, indeed, one gets a fractal distribution 
(although not for all values of $\e$) in small lattices with $L=2,3$.
In this paper, we are mainly interested in the case of large systems $L\gg 1$; thus, we will focus
not on the fractal properties of the field $\{u_k\}$ (which are anyhow hardly accessible in numerics),
but on the statistical properties.

\section{Statistical properties of the KR model on a finite lattice}
\label{sec:kr}

In this section, we report on the scaling properties of the coalescence
process on finite lattices without microdiffusion. This corresponds
to the KR model \eqref{eq:kr} or to the TT model \eqref{eq:tt} with $\e=0$.
In fact, this section aims to extend the scaling relation
\eqref{eq:krdist} (valid for an infinite system) to the case of lattices of finite size $L$. 

In numerical simulations, we start with a uniform initial state $u_k(0)=1$,
$k=1,\ldots,L$. Because in this section masses $u_k(t)$ are integers, 
we refer to these 
quantities as ``number of particles'' at site $k$, or a ``cluster of size $u_k$''.
As expected from the general discussion in Sec. \ref{sec:general}, 
the final state after all the particles merge to
a single site (i.e., they form the maximal cluster of size $L$) 
is $u_k(t)=L\delta_{kj}$, 
where $j(t)$ is the random position of the maximal
cluster. 
The characteristic diffusion time of a particle in the lattice of length $L$ is $T_d(L)=L^2$. 
[Because in the models \eqref{eq:kr} and \eqref{eq:tt} the update is sequential, the time
here and below is measured in units of $L$ to give a possibility 
for every site to move in one effective time unit.]
We thus expect that $T_d(L)$
is the time required for the formation of the maximal cluster.
We now discuss the dynamics in the two relevant regimes.
 
\subsection{Short times: $t\ll T_d$}

\begin{figure}[!htb]
\centering
\includegraphics[width=\columnwidth]{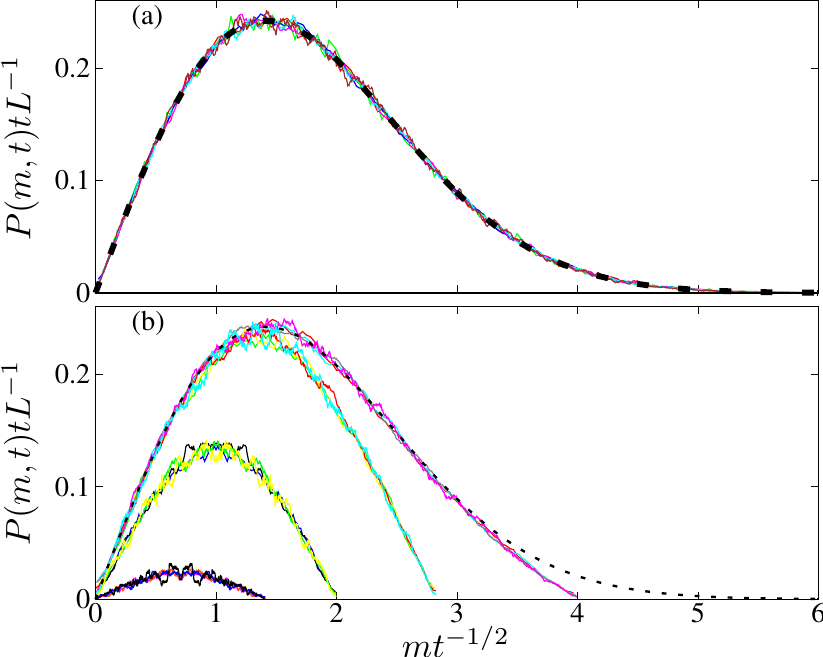}
\caption{Scaled probability distributions for the KR model at different
times and lattice sizes. 
Panel (a): short times, $t\ll T_d$. Data for different $L$ and different times: there are six
nearly overlapping curves for $L=256,\,t=512$;  $L=256,\,t=1024$;
$L=512,\,t=2048$;  $L=512,\,t=4096$; $L=1024,\,t=16384$;  $L=1024,\,t=32768$. The ``guess'' \eqref{eq:fitF} $\mathcal{F}(m/t^{1/2})$ (black dashed line) 
seems also to be very good. In all runs,  
averaging over 32768 realizations is performed. 
Additionally, the curves for $L=512$ and $L=1024$ are 
locally smoothed by a running window (otherwise, fluctuations are relatively large).
Panel (b): the same but for large times. 
For four lattice lengths $L=128,\,256,\,512,\,1024$ the instants of time correspond
to $t=T_d/16,\;T_d/8,\;T_d/4,\;T_d/2$ (nearly overlapping curves from top to bottom);
with $T_d=L^2$.
The curve $\mathcal{F}(x)$  is also shown as the dashed line for comparison on this panel.
}
\label{fig:dst}
\end{figure}

On this time scale, the scaling properties of the infinitely large system should hold.
Indeed,  
KR give a scaling relation \eqref{eq:krdist} (on an infinite lattice) for the average (over realizations of random
advection) number of sites $P(m,t)$ 
possessing a cluster of 
mass $m>0$ at time $t$. We expect this relation to hold on a finite lattice for small times.
To compare results for different lattice sizes $L$, it is convenient to modify the scaling
of \eqref{eq:krdist} by multiplying by $L$ (to pass from a probability to an average number of sites)
and incorporating a factor $m/t^{1/2}$ in the scaling function, so that
\begin{equation}
P(m,t)=t^{-1}L F_1(m t^{-1/2}) \;.
\label{eq:sr}
\end{equation}
Because $L=\sum_m m P(m,t)\approx \int_0^\infty dm\,m P(m,t)$, 
we conclude that normalization of $F_1(x)$ is
independent of $L$:
\[
\int_0^\infty dx\; x  F_1(x)=1 .
\]
It is noteworthy that $\sum_m P(m,t)$ (the total number of non-empty sites) is not normalized to $L$. In fact,
$\sum_m  P(m,t) \simeq L/\sqrt{t}$, which is the equivalent of the relation
$\sum_m c_m(t) \simeq t^{-1/2}$ given by Kang and Redner~\cite{kang1984fluctuation}. 

In Fig.~\ref{fig:dst}(a), we show the simulations for $L=256,512,1024$.   We observe that
(i) The data for different $L$ and $t$ nearly perfectly collapse, and (ii) A guess (black dashed line)
in form of a simple analytical expression
\be
\mathcal{F}(x)=\frac{x}{\sqrt{4\pi}}\exp\left[ -\left(\frac{x}{2}\right)^2\right]
\label{eq:fitF}
\ee
provides a very close fit of the observed data.

\subsection{Large times: $t\lesssim T_d$}

\begin{figure}[!htb]
\centering
\includegraphics[width=0.8\columnwidth]{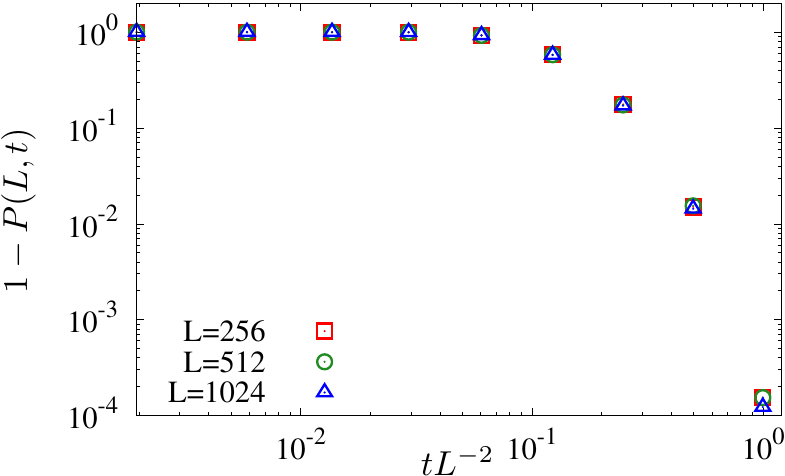}
\caption{Probability for the maximal cluster with occupation $L$ to occur, 
Eq. (\ref{eq:prcl}), as a function
of the scaled time $t/L^2$, for different $L=256,512,1024$.
}
\label{fig:ns}
\end{figure}

For large times $t\approx T_d$, when the probability for the maximal 
cluster to exist $P(L,t)$ is not negligible, 
the distribution 
$P(0<m<L,t)$ deviates from the
one for infinite lattices, as shown in  
Fig.~\ref{fig:dst}(b). Nevertheless, the distributions 
overlap for the same values of $t/T_d$. This 
observation suggests a finite-size generalization of scaling \eqref{eq:sr} 
of the form
\begin{equation}
P(m,t)=t^{-1}L F_2(m t^{-1/2},m/L) +\delta_{m,L}P(L,t)\;,
\label{eq:sr1}
\end{equation}
where in $F_2$, one has $m<L$ (in other words, this function describes all clusters which are less
than the maximal one) and $\delta_{ij}$ is the Kronecker delta. The separation into two parts, the maximal cluster and the rest, allows
for using a continuous approximation for $F_2$.
This scaling function now depends on two arguments, 
$F_2(x,y)$, $x=m/t^{1/2}$, $y=m/L$ and must 
satisfy the following properties: 
(i) As $L\to\infty$, the scaling \eqref{eq:sr} must hold, 
thus $F_2(x,0)=F_1(x)$;
(ii) There is no cluster with a size larger than $L$ 
(and the maximal cluster of size $L$ is not included
to the distribution), thus $F_2(x,y\geq 1)=0$%
~\footnote{The implementation of the latter property
explains why we choose $m/L$ rather than $t/L^2$ as
second argument of $F_2(x,y)$.}.
From the normalization $L=\sum_m m P(m,t)$ it directly follows
the expected scaling for the probability of the maximal cluster:
\begin{equation}
\begin{gathered}
P(L,t)=1-L^{-1}\int_0^\infty dm\,m t^{-1}L F_2\left(\frac{m}{t^{1/2}},\frac{m}{L}\right)\\
=s(t/L^2)\;.
\end{gathered}
\label{eq:prcl}
\end{equation}
This relation is verified in Fig.~\ref{fig:ns}.

\begin{figure}[!htb]
\centering
\includegraphics[width=0.8\columnwidth]{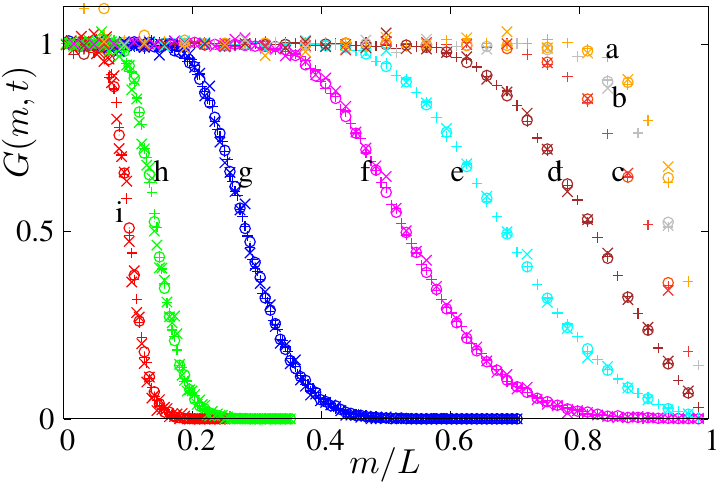}
\caption{Plot of the rescaled distributions  $G(m,t)=t L^{-1}P(m,t)/\mathcal{F}(mt^{-1/2})$ 
at different values of $mt^{-1/2}$ [different colors and grey levels: from top right to bottom left, 
$mt^{-1/2}=4$ (yellow, set a), $mt^{-1/2}=3.46$ (grey, set b), $mt^{-1/2}=2.83$ (dark red, set c), 
$mt^{-1/2}=2$ (brown, set d), 
$mt^{-1/2}=1.41$ (cyan, set e), $mt^{-1/2}=1$ (magenta, set f), $mt^{-1/2}=0.5$ (blue, set g), 
$mt^{-1/2}=0.25$ (green, set f),
$mt^{-1/2}=0.177$ (red, set i)]
as a function of the rescaled variable 
$y=m/L$, for $L=128$ (open circles), $L=256$ (pluses), and $L=512$ (crosses). 
}
\label{fig:cls}
\end{figure}

We check the validity of Eq.~\eqref{eq:sr1} in Fig.~\ref{fig:cls}.
Here, we plot the distributions, rescaled according to 
Eq.~(\ref{eq:sr1}), for several different fixed values of $x=m/t^{1/2}$,
namely we plot $G(m,t)=t L^{-1}P(m,t)/\mathcal{F}(mt^{-1/2})$ versus $y=m/L$. 
For each $x$ considered,  
one can see a nice overlap of data obtained for 
several different sizes $L=128,256,512$, supporting
the Ansatz (\ref{eq:sr1}).
As expected, $G$ tends to 1 for $x\to 0$ and vanishes for $x\to 1$.
Another confirmation of scaling  \eqref{eq:sr1} is the bottom panel of Fig.~\ref{fig:dst}.

\subsection{Roughening properties in KR model}

Let us now describe the mass distribution $u_k(t)$ interpreting it as an ``interface" profile, whose 
width $W(L,t,\e)$ is defined as usual in the following way:
\begin{equation}
W^2=\left\langle \left(u-\langle u\rangle \right)^2\right\rangle,
\label{eq:wdef}
\end{equation}
where $\langle \cdot\rangle$ is both a spatial and a statistical average.
Since this quantity will also be used later on, we made explicit the 
dependence on the parameter $\epsilon$, which
distinguishes the TT model from the KR model (the models
are equivalent for $\e=0$).  

Since the field is conserved, with our choice of the 
initial conditions $\langle u\rangle=1$ always. 
Following the Family-Vicsek scaling approach~\cite{FVscaling}, we can write
\be
W(L,t,0)\sim L^\chi g\left(\frac{t}{L^z}\right),
\ee
where $g(\xi)$ is a suitable scaling function with 
$g(\xi) \sim \xi^\beta$ for $\xi\ll 1$ and $g(\xi)\sim 1$ for $\xi\gg 1$,
so that $W(L,t,0)\sim t^\beta$ at short times ($t\ll L^z$, growth regime) and
$W(L,t,0)\sim L^\chi$ at long times ($t\gg L^z$, saturated regime).
The growth exponent $\beta$, the roughness exponent $\chi$, and the
dynamical exponent $z$ are related by $\chi=\beta z$.

\begin{figure}[!htb]
\centering
\includegraphics[width=0.8\columnwidth]{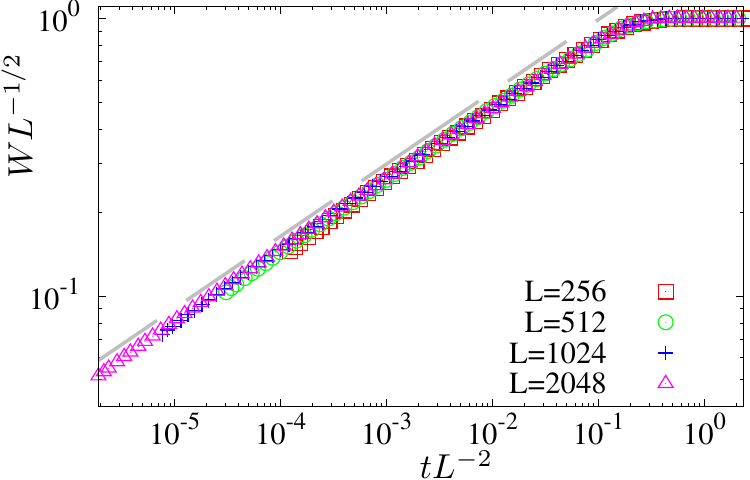}
\caption{Evolution of the width $W(L,t,0)$ in the KR model for different $L$, 
in the rescaled coordinates. 
The gray dashed line has 
slope $0.26$, close to the exponent $\beta$ predicted
by Eq.~(\ref{eq:srkr}). Data collapse on the scaling function $g(t/L^2)$.}
\label{fig:krwidth}
\end{figure}

In the specific case, we are considering in this Section (KR model), 
it is clear that the crossover time between the two regimes is given
by the diffusive time $T_d=L^2$, so that $z=2$.
At short times we can make use of the scaling 
relation \eqref{eq:sr} for the probability to have a height $k$:
\begin{gather*}
L^{-1}P(m,t)=\\
\begin{cases}
t^{-1}F_1(m t^{-1/2}) & m>0\;,\\
1-\int_0^\infty t^{-1}F_1(m t^{-1/2})\,dm=1-\frac{1}{\sqrt{\pi t}} & m=0 \;.
\end{cases}
\end{gather*}
(in the last expression, we calculated the integral using function~\eqref{eq:fitF}).
Thus the width can be computed explicitly using expression \eqref{eq:fitF} for $F_1$:
\begin{gather*}
W^2=\int_0^\infty L^{-1}P(m,t) (m-1)^2\,dm=
\frac{4t^{1/2}}{\sqrt{\pi}}+3.
\end{gather*}
Because the scaling holds for $t\gg 1$, we can assume
\[
W\approx 2\pi^{-1/4} \, t^{1/4},
\]
what yields the exponent $\beta=1/4$ and therefore $\chi=1/2$.
Altogether, we come to the scaling relation
\begin{equation}
W(L,t,0)=L^{1/2} \; g\left(\frac{t}{L^2}\right)\;.
\label{eq:srkr}
\end{equation}
This theoretical prediction is successfully checked in Fig.~\ref{fig:krwidth}.
The scaling function $g(t/L^2)$ could be formally expressed in terms of the scaling
function $F_2(x,y)$, but this would be a futile exercise, especially because
we do not know the analytic form of $F_2(x,y)$ (while we do have a very good expression
for $F_1(x)$!).

At first glance, one may argue that $\beta=1/4$ and $\chi=1/2$ are 
the Edwards-Wilkinson (EW)~\cite{Edwards-Wilkinson-82} roughening exponents in $d=1$.
However, this is just a coincidence because, at variance 
with EW, the distributions are \textit{not} Gaussian.
This should be traced back to the fact that, in the present
model, the total mass is conserved, while in EW, it is not (see, however, a discussion of the conserved EW in Sec.~\ref{sec:smd} below).

In this respect, mass conservation might inspire one to compare the model under consideration with the conserved KPZ equation~\cite{sun1989dynamics,krug1997origins,caballero2018strong}. In one dimension, it reads
\begin{equation}
\begin{gathered}
\frac{\partial \phi}{\partial t}=-\nabla^2[\kappa\nabla^2\phi+\lambda|\nabla\phi|^2]+\eta(x,t),\\
\quad \av{\eta}=0,\\ \av{\eta(x,t)\eta(x',t')}=-2D\nabla^2\delta(x-x')\delta(t-t')\;,
\end{gathered}
\label{eq:ckpz}
\end{equation}
and fulfills mass conservation $\int \phi(x,t)\,dx=const$.
However, the dynamic exponents characterizing 
roughening in this equation  are $z=11/3$, $\beta=1/11$ and $\chi=1/3$ ~\cite{sun1989dynamics},
which are distinctly different from ours.
This is not surprising because
there are two basic differences between \eqref{eq:ade} and \eqref{eq:ckpz}:
(i) In \eqref{eq:ckpz}
the noise is additive and in \eqref{eq:ade} it is multiplicative;
(ii) CKPZ equation is nonlinear while the random advection equation \eqref{eq:ade} is linear.

\section{Evolution to a stationary state in the TT model: overall picture}
\label{sec:ttstat}

\begin{figure}[!htb]
\centering
\includegraphics[width=\columnwidth]{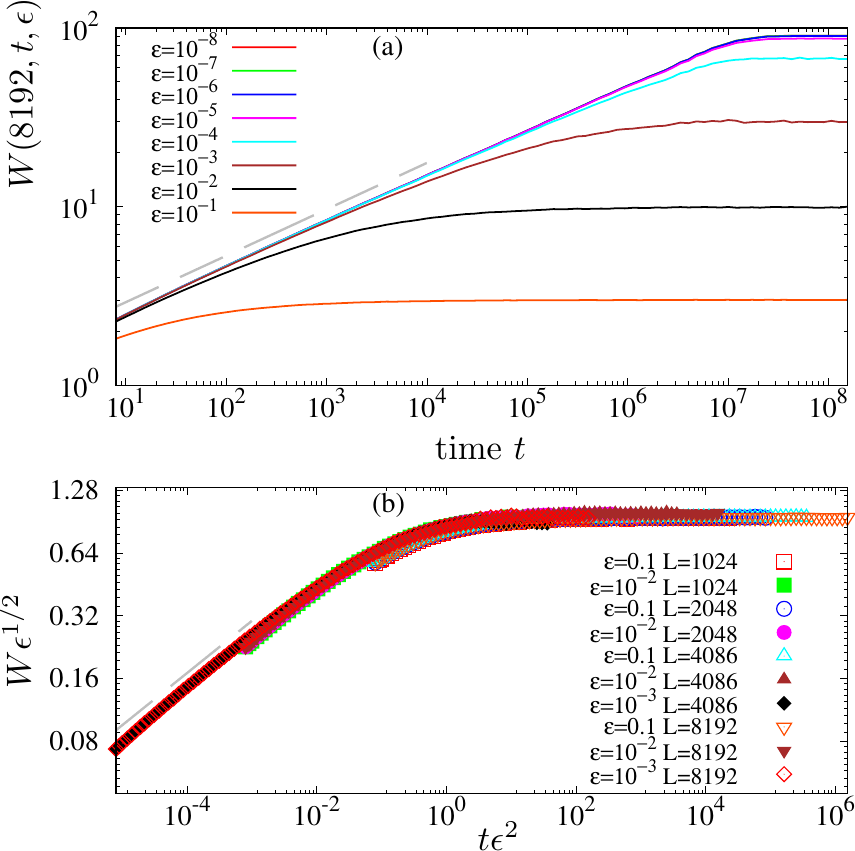}
\caption{Panel (a): field widths $W(L,t,\epsilon)$ as a function of time, 
for different $\e$ (increasing from top to bottom) and $L=8192$.
The dashed line has a slope of $0.26$, which is close to the theoretical value
$0.25$ derived for $\epsilon=0$.
Panel (b): Widths of the field in dependence on time for different $\e,L$ in 
scaled coordinates. The dashed line has a slope of $0.25$.
}
\label{fig:ttw}
\end{figure}

Let us now turn to the case where parameter $\epsilon$ in \eqref{eq:tt} is non-zero.  In other words, both macrodiffusion and microdiffusion are present. We first performed the same roughening experiment as above but with the TT model. The results for different values of $L$ and $\e$ are presented in Fig.~\ref{fig:ttw}. Both panels clearly indicate that, for a finite system, a statistically stationary regime establishes at long times.  However, the evolution of the width crucially depends on parameter values $\e,L$. In panel (a)  of Fig.~\ref{fig:ttw}, where we fix $L=8192$,  one can clearly see that the values of $\e$ can be separated into two ranges. For  $\e\lesssim 10^{-5}$, there is no significant $\e$-dependence in $W(L,t,\epsilon)$; the curves follow the roughening in the KR model \eqref{eq:srkr}. In contradistinction, for $\e\gtrsim 10^{-4}$, the dependence of $W(L,t,\epsilon)$ on $\e$ is significant, and the saturated width $W(L,\infty,\epsilon)$ decreases with $\e$. Therefore, for large values of $\e$, we plot the data in another scaling that includes $\e$, in Fig.~\ref{fig:ttw}(b). Now the data for different $L$ overlap, which indicates that the ``roughening'' is not $L$-dependent. In other words, in this regime, there is no true roughening because the saturated width is system size-independent (but depends on the parameter $\e$).

\begin{figure}[!htb]
\centering
\includegraphics[width=0.9\columnwidth]{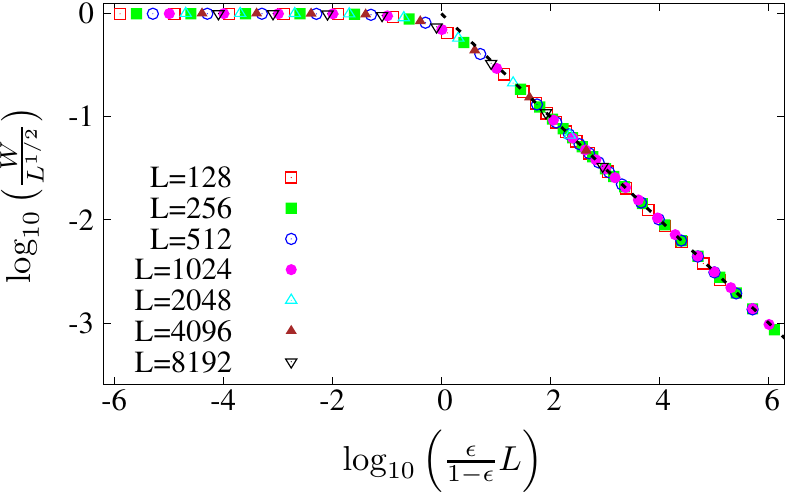}
\caption{Asymptotic ($t\to\infty$) roughness $W(L,\infty,\e)$ as a function of the system size
$L$ and of the parameter $\e$. Data collapse shows that
$W(L,\infty,\e) = L^{1/2} f(L\e/(1-\e))$, where the scaling function $f(u)$
has the limiting behavior $f(u) = 1$ for $u\ll 1$ and $f(u) = u^{-1/2}$
for $u\gg 1$.  The dashed line has a slope of $-1/2$.\\
}
\label{fig:stw}
\end{figure}

A clean way to analyze the separate effects of the system size $L$ and of the parameter $\e$ is 
to focus on the asymptotic, time-independent width $W(L,\infty,\e)$.
This is done in Fig.~\ref{fig:stw} where we plot $W(L,\infty,\e)/L^{1/2}$ 
\textit{versus} $\e L/(1-\e)$.
The data of Fig.~\ref{fig:stw} cover a wide range of values of $L$ and $\e$ and 
clearly indicate the existence of two types of stationary states:
\begin{itemize}
\item \textbf{Macrodiffusion-dominated regime}. 
This regime corresponds to the leftmost part
of the graph where the scaled width $WL^{-1/2}$ does not depend on $\e$. Here the width
scales $W\sim L^{1/2}$ like in the KR model at $\e=0$. Nevertheless, the state here
is nontrivial and will be discussed in detail in Section~\ref{sec:sme} below.
\item \textbf{Microdiffusion-dominated regime}. This regime corresponds to the right-most
part of the curves in Fig.~\ref{fig:stw}, where the scaling $WL^{-1/2}\sim \left(\frac{\e}{1-\e} L\right)^{-1/2}$
holds. This means that here the width does not depend on the system size $L$: $W\sim 
\left(\frac{\e}{1-\e}\right)^{-1/2}$. We discuss this regime in Section \ref{sec:lae} below.
\end{itemize}
The crossover between two regimes occurs at $\left(\frac{\e}{1-\e}L\right)\approx 1$, i.e. 
at $\e L\approx 1$. We remark that the microdiffusion-dominated regime is attained
when $\e \gg 1/L$ but also for $\e\to 1$ and rather small $L$.
A final comment about simulations is in order here.
While it is relatively easy to vary  parameter $\e$
in a wide range, for the length $L$, we can hardly significantly increase the range beyond 
several thousand.

Below in Sections \ref{sec:lae} and \ref{sec:sme}, we will focus on the 
detailed analysis of stationary regimes for microdiffusion- and macrodiffusion-dominated regimes, respectively.

\section{Strong microdiffusion $\e >L^{-1}$}
\label{sec:lae}
 
The discussion above shows that the system length is irrelevant here.

\subsection{Mean-field theory}
\label{sec:mft}
We start with the mean-field theory, where spatial correlations are neglected 
(our approach is similar to that of Ref \cite{takayasu1993non} but does not 
coincide with it).  With probability $1/2$, 
each site either delivers part of its field to a neighbor,
or receives a part of the neighbor's field. Thus, the updating rule 
for a field  at a given site reads $u\to \bar{u}$,
\begin{equation}
\bar{u}=\begin{cases} \e u& \text{ Prob }1/2\;,\\
u+(1-\e)v&\text{ Prob }1/2\;,
\end{cases}
\label{eq:upd}
\end{equation}
where $v$ is the field of the neighbor. In the mean-field approach, we assume statistical independence
of $u$ and $v$, which have the same distribution. This allows for expressing the evolution
of the density through a Perron-Frobenius operator ($w$ and $\bar{w}$ denote densities at the subsequent time steps)
\begin{equation}
\begin{gathered}
\bar{w}(x)=\av{\delta(x-\bar{u})}=
\frac{1}{2}\int_0^\infty du\, w(u)\delta(x-\e u)+\\
\frac{1}{2}\iint_0^\infty 
du\,dv\; w(u) w(v)\delta(x-u-(1-\e)v)=\\
=\frac{1}{2\e} w\left(\frac{x}{\e}\right)
+\frac{1}{2}\int^{x(1-\e)^{-1}}_0 dv\,w\left(x-(1-\e)v)\right) w(v)\;.
\end{gathered}
\label{eq:fpd}
\end{equation}
Unfortunately, we cannot solve this equation analytically, 
except for the case $\e=1/2$, for which one can easily check that the 
solution is an exponential distribution $w(u)=\exp(-u)$.
Indeed, in this case the calculation of the r.h.s. of \eqref{eq:fpd} is straightforward:
\begin{gather*}
w(2x)+\frac{1}{2}\int_0^{2x} dv\; e^{-x+v/2} e^{-v}=\\
=e^{-2x}+\frac{e^{-x}}{2}\int_0^{2x}e^{-v/2}dv=e^{-x}\;.
\end{gather*}

It is however possible to express, for arbitrary $\e$, all the moments $M_n=\av{u^n}$
explicitly in a recursive manner. Indeed, directly from \eqref{eq:upd} it follows that
$M_n=\frac{1}{2}\av{(\e u)^n}+\frac{1}{2}\av{(u+(1-\e)v)^n}$, giving
\begin{equation}
M_n=\frac{\sum_{k=1}^{n-1}\binom{n}{k}(1-\e)^k M_{n-k}M_k}{1-\e^n-(1-\e)^n}\;.
\end{equation}
Since the total field is conserved, the value of $M_1$ is arbitrary, and we 
may take $M_1=1$, 
as enforced in the numerical simulations. This yields $\av{u^2}=1/\e$, so that the mean width defined
according to \eqref{eq:wdef} is $W^2=M_2-M_1^2=\frac{1-\e}{\e}$.
Figure~\ref{fig:stw} proves that this result is correct in the large $\e L$ regime.

 \begin{figure}[!htb]
 \centering
 \includegraphics[width=\columnwidth]{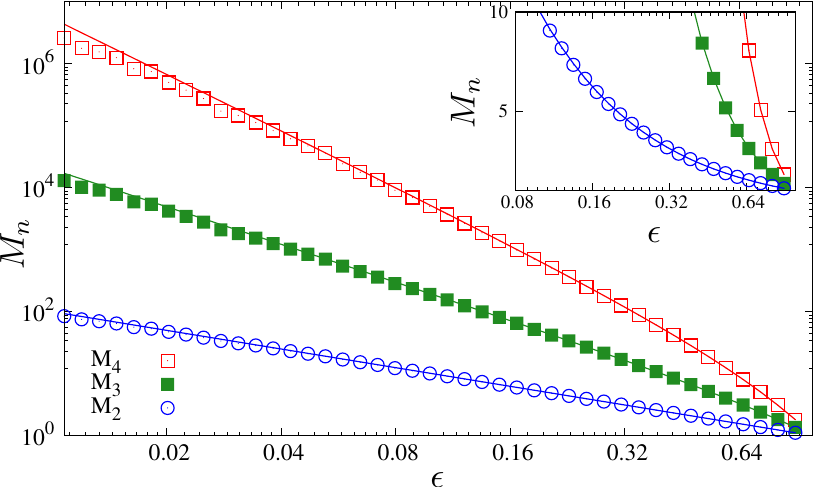}
 \caption{Comparing numerically found moments 2-4 (markers, in a lattice of $L=1024$) 
 with theoretical formulae (lines). The inset shows a region of large $\e$ with a linear scale of $M_n$-axis.
 }
 \label{fig:m24} 
 \end{figure}

\begin{figure}[!htb]
 \centering
 \includegraphics[width=0.85\columnwidth]{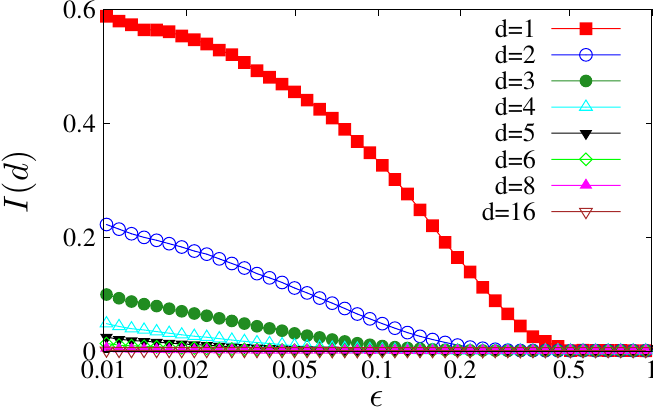}
 \caption{Mutual information for different distances $d$ between the sites.
 Details of calculations: $L=1024$, number of patterns in statistical averaging $1024$, 
 number of bins $64$ (bins are spaced so that all bins have the same probability $1/64$).
 }
 \label{fig:me} 
 \end{figure}

To test more thoroughly the accuracy of the approximations, in Fig.~\ref{fig:m24}
we compare the mean-field values of the first three nontrivial 
moments with their numerical values. 
While the comparison seems to support
the main assumption that the neighboring sites are statistically independent, an additional check for this dependence gives a different picture. 
For a quantitative characterization of the independence of two distributions 
 $w(u_n)$ and $w(u_{n+d})$ at sites separated by distance $d$ we uses the mutual information 
 \[
 I(d)=\sum_{i,j} W_{ij}\log\frac{W_{ij}}{p_i q_j}\;.
 \]
where $p_i$ and $q_j$ are binned probabilities, and $W_{ij}$ is the joint binned probability. For independent random variables, mutual information vanishes, but in real calculations, it is always positive. The values of $I(d)$ for large distances $d$ serve as  ``surrogates'', giving
the numerical level of mutual information for practically
independent distributions.
The results (Fig.~\ref{fig:me}) suggest that the independence of neighbors might be
\textit{exact} for $\e\geq 0.5$. But, instead, mutual dependence turns on
for $\e<0.5$. On the other hand, even for $\e=0.01$, only some five neighboring sites
are interdependent according to the mutual information criterion.

\subsection{Limit of strong microdiffision $\e\to 1$}
\label{sec:smd}

Let us rewrite the local TT model as an application of a matrix 
\[
\mathsf{A}(\mu,\pm)=\begin{cases}\begin{pmatrix} 1-\mu& 0\\ \mu &1\end{pmatrix} & \text{ prob } 1/2\;,\\
\begin{pmatrix} 1& \mu\\ 0 &1-\mu\end{pmatrix} & \text{ prob } 1/2\;.
\end{cases}
\]
where a portion $\mu=1-\e$ is moved to the right (to the left). 
To have a symmetric situation, suppose that this matrix is applied twice (thus, one has four combinations). 
Furthermore, we assume $\mu\ll 1$, in this case $\mathsf{A}(\mu,+)\mathsf{A}(\mu,-)\approx
\mathsf{A}(\mu,-)\mathsf{A}(\mu,+)$.
Then in the 1st order in $\mu$
\[
\mathsf{A}^2=\begin{cases}\begin{pmatrix} 1-\mu& \mu\\ \mu &1-\mu\end{pmatrix}
=I+\mu \begin{pmatrix} -1& 1\\ 1 &-1\end{pmatrix} & \text{ prob } 1/2\;,\\
\begin{pmatrix} 1-2\mu& 0\\ 2\mu &1\end{pmatrix}=I+2\mu \begin{pmatrix} -1& 0\\ 1 &0\end{pmatrix} & \text{ prob } 1/4\;,\\
\begin{pmatrix} 1& 2\mu\\ 0 &1-2\mu\end{pmatrix}=I+2\mu \begin{pmatrix} 0& 1\\ 0 &-1\end{pmatrix} & \text{ prob } 1/4\;.
\end{cases}
\]
(here, $I$ is the unit matrix).
To obtain a continuous in space formulation, we attribute operator $\partial_{xx}$ to the first matrix, 
and operators $\pm \partial_x$
to the second and the third matrices. In this way, we approximate the evolution with
\[
\partial_t u(x,t)=\mu \partial_x(V u)+\mu\partial_{xx}u,\qquad V=\pm \frac{\Delta x}{\Delta t}\;.
\]
Because $V$ has independent values at different sites and different time steps, 
we can model velocity with a $\delta$-correlated noise field
\begin{gather*}
\partial_t u(x,t)=\mu \partial_x(\xi(x,t) u)+\mu\partial_{xx}u,\\
\av{\xi(x,t)\xi(x',t')}=\delta(x-x)\delta(t-t')\;.
\end{gather*}
Rescaling time $\mu t=\tau$ we obtain
\begin{gather*}
\partial_\tau u(x,\tau)=\mu^{1/2} \partial_x(\eta(x,\tau) u)+\partial_{xx}u,\\ \av{\eta(x,\tau)\eta(x',\tau')}=
\delta(x-x)\delta(\tau-\tau')\;.
\end{gather*}

Let us suppose that $u=u_0+\mu^{1/2}u_1+\mu u_2\ldots$. 
Substituting this, we get in the leading order
\[
\partial_\tau u_0=\partial_{xx}u_0\;,
\]
which yields a uniform asymptotic state $u_0=const$. We suppose $u_0=1$, like in the lattice model above.

In the next order, we get
\[
\partial_\tau u_1=\partial_x (\xi(x,\tau))+\partial_{xx}u_1
\]
which is the conserved version of the Edwards-Wilkinson (EW) equation~\cite{Edwards-Wilkinson-82}.

Smith et al. ~\cite{Smith_etal-17}  considered this EW equation and, in particular, demonstrated
that the variance diverges (UV catastrophe). They did not perform a cutoff at the lattice size, but from their Eq.~(8)
it follows that $\text{var}(u_1)\approx \ell^{-1}$ where $\ell$ is the lattice spacing. The total variance
is $\text{var}(u)\approx \mu \ell^{-1}$, in agreement with the result for the lattice model. Furthermore, from Gaussianity of $\xi(x,t)$
it follows that the distribution of $u_1$ is Gaussian. We show that the field $u$  in the lattice TT model is indeed 
Gaussian in the limit $\e\to 1$ in Appendix~\ref{app:gauss}. Thus, we conclude that the limit $\e\to 1$ corresponds
to the conserved version of the Edwards-Wilkinson stochastic differential equation.

\subsection{Field distribution}

\begin{figure}[!htb]
 \centering
 \includegraphics[width=\columnwidth]{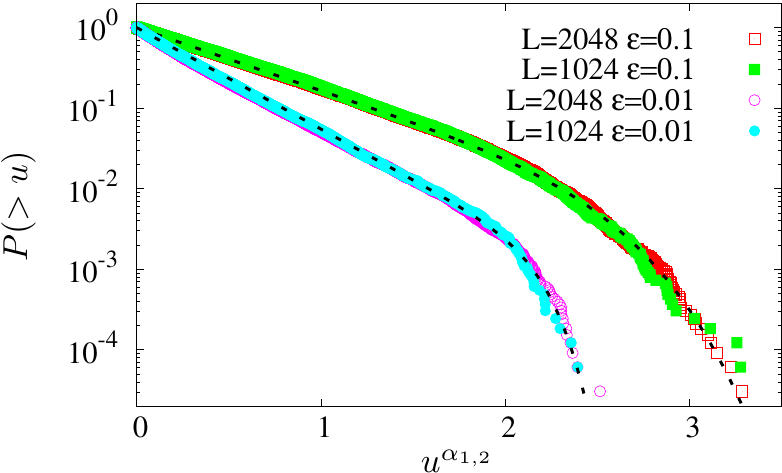}
 \caption{Cumulative field distributions for $\e=0.1$  and 
 $\e=0.01$, and different $L$. Markers are the simulation data, 
 dotted lines are fits with expression (\ref{eq:stre}).
 For the exponents $\alpha_{1,2}$
 the theoretical expression \eqref{eq:alpha} is used.
 }
 \label{fig:stre} 
 \end{figure}

As mentioned above, we can solve Eq.~\eqref{eq:fpd} for the field distribution only in a special case $\e=1/2$. 
Numerical simulations have shown that for $\e<1/2$, the distribution has a power-law singularity at $u\to 0$, and 
cumulative distribution can be well approximated by a stretched exponential with a Gaussian cutoff:  
\begin{equation}
P(>u)=\exp[-A (uL)^\alpha -B (uL)^2]\;.
\label{eq:stre}
\end{equation}
We illustrate this in Fig.~\ref{fig:stre}, where we show in rescaled
coordinates the distributions for $\e=0.1$ and $\e=0.01$. 

\begin{figure}[!htb]
\centering
\includegraphics[width=0.85\columnwidth]{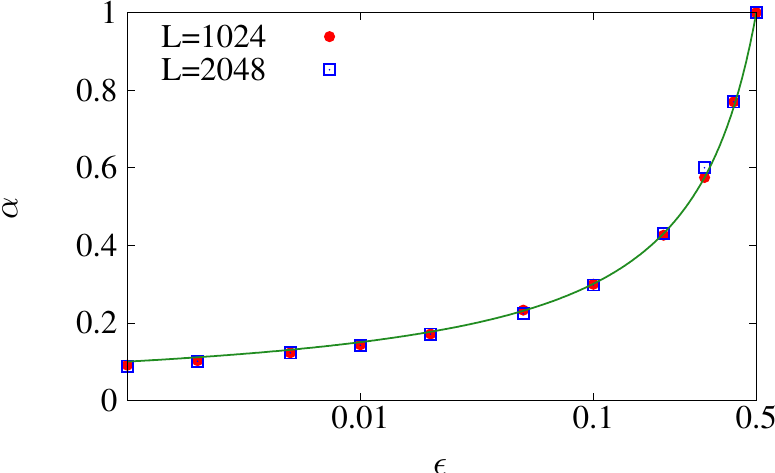}
\caption{Numerical values of 
the exponent $\alpha$ of the stretched-exponential 
part of the cumulative distribution (\ref{eq:stre}) as a function of $\e$
(markers) together with the analytic estimate, Eq. (\ref{eq:alpha}).
}
\label{fig:na}
\end{figure}

Although we cannot derive expression \eqref{eq:stre},
we can estimate the exponent $\alpha$ assuming the validity of \eqref{eq:stre}.
Let us suppose that the cumulative distribution 
$P(>u)=\int_u^\infty w(y)\,dy$ has the form of a stretched exponential
$P(>u)=\exp[-a u^{\alpha}]$ for small $u$ and $\alpha<1$. 
Then, the density has a power law singularity at small $u$ of the form
 \[
 w(u)=a \alpha u^{\alpha-1}\exp[-a u^\alpha]\approx A u^{\alpha-1}.
 \]
 Let us look which value of $\alpha$ is consistent with the Perron-Frobenius
 equation \eqref{eq:fpd}. Substituting, we get
 \begin{gather*}
 A u^{\alpha-1}\approx\\
  \frac{1}{2}\frac{1}{\e}A \frac{u^{\alpha-1}}{\e^{\alpha-1}}+
 \frac{1}{2}\int_0^{u/(1-e)} A^2 (u-(1-\e)v)^{\alpha-1}v^{\alpha-1})\;dv=\\ 
 \frac{A u^{\alpha-1}}{2\e^{\alpha}}+\frac{A^2}{2(1-\e)^\alpha}u^{2\alpha-1}
 \int_0^1 (1-z)^{\alpha-1}z^{\alpha-1}\, dz\;.
 \end{gather*}
 Neglecting the last term, we obtain the consistency condition $2\e^\alpha=1$ which means
 \begin{equation}
 \alpha=-\frac{\log 2}{\log\e}\;.
 \label{eq:alpha}
 \end{equation}
 This equation is in excellent agreement with the numerics 
 as shown in Fig. \eqref{fig:na}.
  
\subsection{Time correlations}

\begin{figure}[!htb]
\centering
\includegraphics[width=0.9\columnwidth]{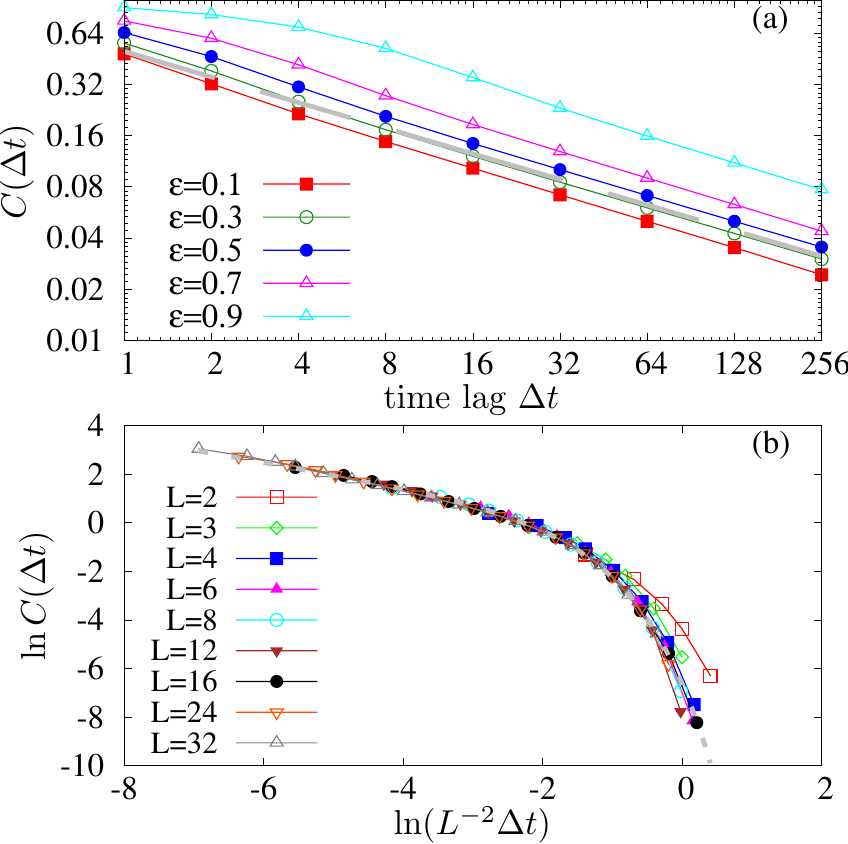}
\caption{Panel (a): Time-correlations 
functions for a lattice of length $L=1024$ and different $\e$. The dashed grey line on this panel has a slope of $-1/2$.
Panel (b): Time-correlations in scaling coordinates for different $L$ and $\e=0.6$.
The dashed gray line is a fit $\log y=-0.511-0.5\log x-6.122 \exp(x)$.
}
\label{fig:tcor}
\end{figure}

In this section, we discuss the one-site temporal correlation function of the field
(remember that we set $\langle u\rangle=1$)
$$
C(\Delta t)=\av{(u_n(t)-1)(u_n(t+\Delta t)-1)}.
$$
The calculated time-correlation 
function is shown in panel (a) of  Fig \ref{fig:tcor}. It appears that the 
correlations decay as a power law 
$(\Delta t)^{-1/2}$. In contrast, for $L=2$ the decay 
is exponential (see Appendix \ref{sec:bc} and 
the  panel (b) of
Fig \ref{fig:tcor}). Thus, one can expect a crossover at small $L$, we show in Fig \ref{fig:tcor}(b) 
how correlations for a fixed $\e=0.6$
depend on $L$. Here scaled coordinates are used; one
can see that starting from $L=4$, the scaling law 
$C(\Delta t)\approx L^{-1} b(\Delta t L^2)^{-1/2}\exp[-a \Delta t L^2]$ works well 
(fitted values for $a,b$ are in the caption).

\section{Weak microdiffusion $\e < L^{-1}$}
\label{sec:sme}

\begin{figure}[!htb]
\centering
\includegraphics[width=\columnwidth]{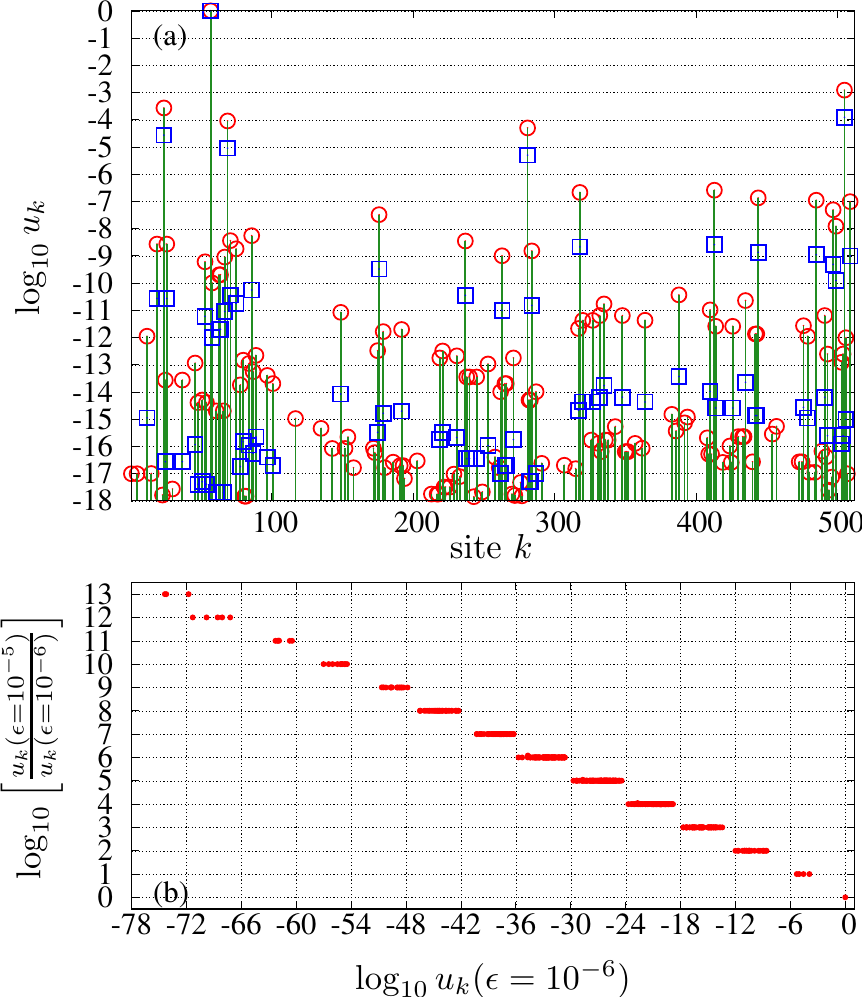}
\caption{Panel (a): Snapshots of the fields with $\e=10^{-5}$ (red circles) and $\e=10^{-6}$ (blue squares)
for $L=512$. Green vertical lines show the positions of the main peaks (i.e., those with markers; there are many other peaks with
masses smaller than $10^{-18}$, cf. bottom panel).
Panel (b): The logarithmic difference in the levels of the peaks (i.e., the distance between red circles and blue squares
of panel (a) at each peak) vs. the level of the blue squares.
}
\label{fig:ttf512}
\end{figure}

\subsection{Hierarchical structure of peaks}

We start our treatment of the case of very small microdiffusion
with a visualization in Fig. \ref{fig:ttf512} (a) of a snapshot of a field $\{u_k\}$ in a statistically stationary
regime (i.e., at times larger than characteristic transient time $L^2$), at small values of 
microdiffusion parameter $\e$.
At $\e=0$, the field is just one peak (the maximal cluster)
at which the whole initial ``mass'' is concentrated, 
at a random spatial position.
For better comparison of fields for different sizes of the lattice $L$,
we use below in this section the normalization $\sum_k u_k=1$; thus, the single peak
for $\e=0$ has mass one. Together with this maximal cluster, one observes in Fig.~\ref{fig:ttf512}(a) peaks at different
levels with a strong separation (several orders) between them.

To qualitatively understand this hierarchical structure of the field (which we will quantitatively characterize below),
let us start with a single peak at $\e=0$ and switch to a finite but 
small value of $\e$. Then, the randomly moving main peak will leave behind secondary peaks of mass 
$\approx \e$. These peaks will also move, leaving the next generation of peaks of mass $\approx \e^2$;
they can also merge and be absorbed by the main peak 
(the size of which remains close to one -- this is where the condition $\e L<1$ plays its role). Thus, one can expect peaks at levels 
$\sim\e,\sim\e^2,\sim\e^3,\ldots$. However, this hierarchy is not very distinct, although 
recognizable, in the single profile at fixed $\e$; see the set of red circles or
of blue squares in
Fig. \ref{fig:ttf512}(a). To separate different levels in a more apparent way,
we perform a simultaneous run of the TT model at two different values of parameter $\e$:
$\e_1$ and $\e_2$.
This means that the same random choices for advection steps \eqref{eq:tt} are chosen in two runs. 
As a result, the peaks in the two runs coincide in position but differ in their 
height by factor $(\e_1/\e_2)^m$ with integer $m$. 
For an illustration in Fig.~\ref{fig:ttf512}(a)
we have chosen $L=512$, $\e_1=10^{-5}$ (red circles)
and $\e_2=10^{-6}$ (blue squares). The grid in the $y$-axis corresponds to 
the ratio $\e_1/\e_2=10$. One can see that the main peaks in the two runs coincide. There are
four peaks at the next level, with separation between them by factor $\e_1/\e_2=10$. The 
number of peaks at the next level is larger; there, the separation is $(\e_1/\e_2)^2=100$, 
etc. To make the correspondence of the separation and the level apparent,
we plot in Fig.~\ref{fig:ttf512}(b) all the $L=512$ values of $\{u_k\}$ from the snapshot
Fig.~\ref{fig:ttf512}(a) in the coordinates ``mass vs separation'' 
(both axes logarithmic). One can see that the separations are very well ``discretized''
at integers of $\log_{10}(\e_1/\e_2)$ ($y$-axis), while the levels in the field values 
are spread much wider ($x$-axis), and these widths for deep levels are of the same order as 
$-\log_{10}\e_2=6$.

\begin{figure}[!htb]
\centering
\includegraphics[width=\columnwidth]{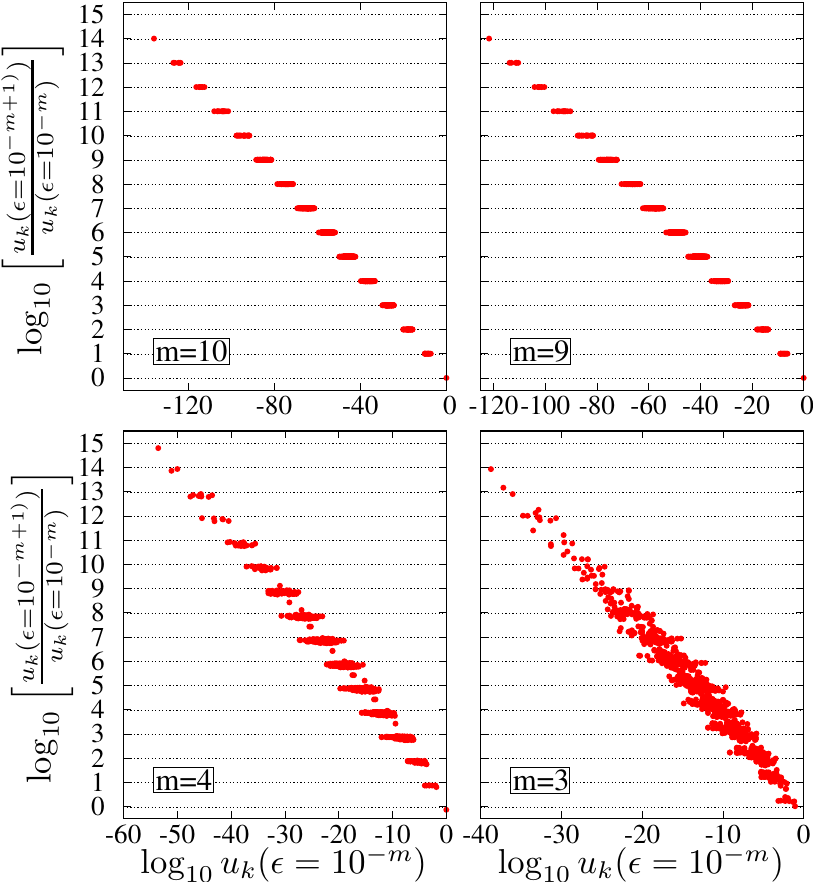}
\caption{The same plots as in Fig \ref{fig:ttf512}(b), but for different values of $\e=10^{-m}$. One can see that for large
$\e$, the order structure becomes blurry  and eventually, no clear separation is seen for large $\e$.}
\label{fig:ttf1024}
\end{figure}

To illustrate that this hierarchical structure appears for small enough $\e$ only,
we show in Fig.~\ref{fig:ttf1024} the same plots as Fig.~\ref{fig:ttf512}(b) for $L=1024$ and different values
of $\e_1=10^{-m}$ and $\e_2=10 \e_1$. One can see that the separation is apparent for $m=10,9$
but becomes less distinct for $m=4$ and is practically not seen for large $\e$ ($m=3$).

\subsection{Order kinetic model}
\label{sec:km}

In this Section, we present an effective model (termed ``order kinetic model'' (OKM)) to
describe the structure observed in the simulation
for small $\e$, Figs.~\ref{fig:ttf512}-\ref{fig:ttf1024}. Motivated by the observed hierarchical structure, 
we attribute to each site $k$ an integer-valued order $\mu_k$ and assume that the field $u_k$ is 
represented as 
\begin{equation}
u_k \approx \rho \epsilon^{\mu_k}\;.
\label{eq:km1}
\end{equation}
Here $\mu_k\geq 0$ is an integer called ``order'' of the field at site $k$.
Parameter $\rho\approx 1$ is a
normalization factor (it roughly corresponds to the mass of 
the (unique) site having zero order in the asymptotic
steady state). In the following, we will refer to
the sites having the same value of $\mu$ as the 
peaks of order $\mu$ and denote by $n(\mu)$ their 
number and by $p(\mu)=n(\mu)/L$ their fraction (i.e., the probability to observe specific order).
The expression \eqref{eq:km1} corresponds to an approximation in which the horizontal steps in Fig.~\ref{fig:ttf512}(b) have 
zero width and zero height.

Let us now rewrite the update rule \eqref{eq:tt} in terms of the orders
$\mu_k$, by considering 
two neighboring sites $(i,j)$ corresponding respectively to orders $(\mu_i,\mu_j)$, with
the direction of advection $i\to j$.
Since the mass $(1-\e)u_i$ is transferred to site $j$ and the mass $\e u_i$ is left to site $i$,
we write the following update rule for the orders:
\begin{equation}
\begin{aligned}
\mu_i&\to \mu_i+1\;,\\
\mu_j &\to \mbox{MIN}\{\mu_i,\mu_j\}\; .
\end{aligned}
\label{eq:km}
\end{equation}
The approximation here follows from our perfect discretization of the levels: we neglect changes in the
field, if the addition is smaller than the existing field by factor $\e^{m}$, $m\geq 0$.

Special care should be taken about the sites with minimal possible order $\mu_{min}=0$.
As it follows from \eqref{eq:km}, the number of such sites can only decrease, and eventually, there is only one such site in the lattice. With one site having zero order, 
this situation is an absorbing
state in model \eqref{eq:km}.

Before proceeding, we notice that the OKM \eqref{eq:km} is, in fact, a skew (unidirectionally coupled) system:
Higher-order peaks do not influence the zero-order peak; the first-order peaks interact only with each other (can coalesce) and with the zero-order peak (can be ``emitted'' or ``absorbed'' by it), etc. We will use 
this property in section \ref{sec:dlop} below.

\subsection{Mean field approach}

\begin{figure}[!htb]
\centering
\includegraphics[width=\columnwidth]{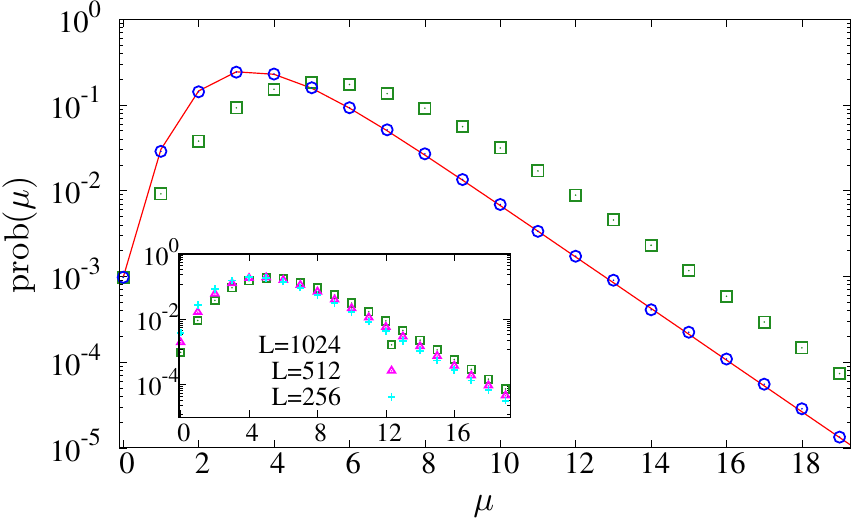}
\caption{Comparison of a theoretically obtained distribution of peaks $p(\mu)$ (Eq. \eqref{eq:rec2}, red curve)
with numerics for the TT model (green squares) and for the global coupled TT model \eqref{eq:ttl}, blue circles. 
All these data are for $L=1024$. In the inset, we
also show data for the TT model with $L=512$ and $L=256$.}
\label{fig:KMTaTa}
\end{figure}

In the framework of the OKM, one can apply the same mean-field approach
as in Section \ref{sec:mft} to write an equation for the evolution of the probabilities $p(\mu)$. 
The basic assumption is the independence of
neighboring values of $\mu_{i,j}$ in \eqref{eq:km}. Thus, the minimum MIN in~\eqref{eq:km} should be calculated
as a minimum value of two independent random variables having the same distribution $p(\mu)$:
\begin{gather*}
\text{prob}(\mbox{MIN}(\mu_i,\mu_j)>\mu)=\text{prob}(\mu_i>\mu)\cdot\text{prob}(\mu_j>\mu)=\\
(1-\text{prob}(\mu_i\leq \mu))\cdot(1-\text{prob}(\mu_j\leq\mu))\;.
\end{gather*}
This leads
to the following expression for this distribution, valid for $\mu\geq 1$  (this expression is analogous to \eqref{eq:fpd}):
\begin{equation}
p(\mu)=\frac{1}{2}p(\mu-1)+\frac{1}{2}\left[(1-\sum_{\nu=0}^{\mu-1}p(\nu))^2 - (1-\sum_{\nu=0}^{\mu}p(\nu))^2\right]\,.
\label{eq:rec1}
\end{equation}
The first term corresponds to the case where the site is a ``source'' (this happens with probability $1/2$);
the second term corresponds to the case (also with probability $1/2$) when the site is a ``destination''.
One can see from \eqref{eq:rec1}, that $p(\mu)$ depends only on values $p(\nu)$ with $\nu<\mu$. In terms of $p(\mu$),
expression \eqref{eq:rec1} is a quadratic equation.
Its solution begets a recursion relation
\begin{equation}
\begin{gathered}
p(\mu)=\sqrt{s^2(\mu-1)+p(\mu-1)}-s(\mu-1)\;,\\ s(\mu)=\sum_{\nu=0}^{\mu}p(\nu)\;,
\end{gathered}
\label{eq:rec2}
\end{equation}
which has to be iterated started from $p(0)=1/L$. We compare this solution with
numerically obtained distribution in Figure \ref{fig:KMTaTa}. One can see that correspondence is
not good, what indicates that in the TT model \eqref{eq:tt} 
the correlations between the neighboring sites are large and cannot be neglected.
On the other hand, for the global coupled version of the TT model \eqref{eq:ttl}, where correlations 
are expected to be very small in the thermodynamic limit, the correspondence is very good.

\subsection{Dynamics of lowest-order peaks}
\label{sec:dlop}

\begin{figure}[!htb]
\centering
\includegraphics[width=\columnwidth]{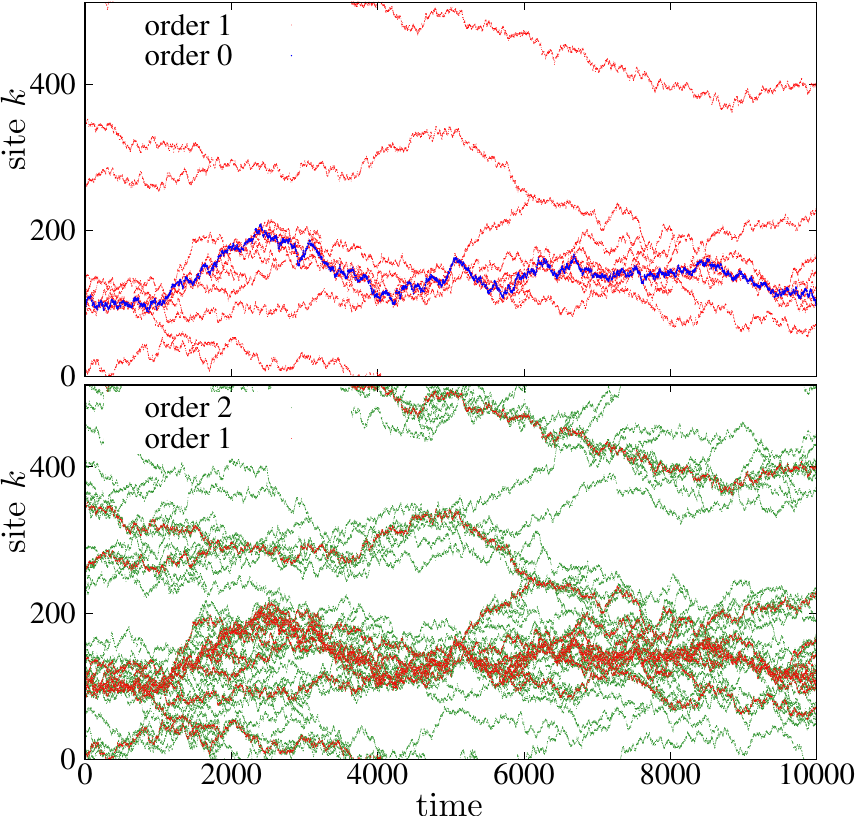}
\caption{Trajectories of peaks of orders 0 (blue, dark) and 1 (red, middle dark) (top panel),
and of orders 1 (red, middle dark) and 2 (green, light grey) (bottom panel) (simulations of the TT model \eqref{eq:tt} 
for $L=512,\;\e=10^{-6}$).
Simulations were started from random initial conditions but an initial transient of duration 
$L^2$ was dismissed.
}
\label{fig:2df512}
\end{figure}

As demonstrated above,
for the OKM, the mean-field approximation is not very successful because
of correlations in the peak positions. Such a correlation is not very surprising
because the first-order peaks are ``daughters'' of the zero-order peak and thus are located close to it;
the same holds for other orders (``The apple never falls far from the tree'').
To get insight, we visualize the dynamics of the peaks of
orders $\mu=0,1,2$.
In Fig.~\ref{fig:2df512}, we compare the trajectories of the main (zeroth order)
peak (blue) with the ones of peaks of order one (red) 
and two (green). One can clearly see that the first-order peaks are mainly in
the vicinity of the zero-order peak (from which they are created), but some leave this vicinity, diffuse and merge.
A similar creation-merging is observed at the bottom panel, where peaks of order one and 
two are depicted.

\begin{figure}[!htb]
\centering
\includegraphics[width=0.9\columnwidth]{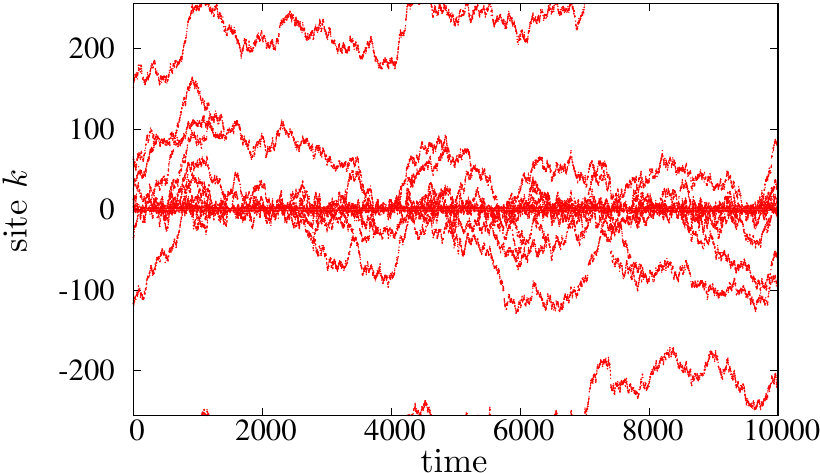}
\caption{Trajectories of the distances of the peaks of  levels one from the main peak  
for $L=512$ (thus, the position of the main peak is at zero) in the TT model \eqref{eq:tt}.}
\label{fig:2df-rel}
\end{figure}

Let us now describe the relative motion of the main peak and the 1st order peaks. To this aim, 
let us focus on Figure \ref{fig:2df512} (middle panel), which shows the main peak (order 0)
and the peaks of order 1. Let us redraw this figure by
plotting the differences in the positions of the peaks of order one and the position of order 
zero; see Fig. \ref{fig:2df-rel}(in fact, here, another random realization is taken).

To describe Fig. \ref{fig:2df-rel}, we can formulate a reduced version of the OKM \eqref{eq:km}, 
which takes into account
only peaks of the 1st order (we denote their masses as $w_k$) and the main one (zero-order peak).
Moreover, it is instructive to go beyond the OKF and distinguish masses of the 1st-order peaks (although
later, we will ignore them).
Because the motions of the main peak and of other sites are independent,
we place the main peak at zero and fix it. The dynamics of all other sites follows the  KR model (TT model with $\e=0$). Namely, at time $t$, a pair of points $k,l$ is chosen randomly, with $l=k\pm 1$.
If $k=0$, then we set $w_l(t+1)=w_l(t)+1$. If $k\neq 0$ and $l\neq 0$, then the dynamics is as follows
\[
w_k(t+1)=0,\qquad w_l(t+1)=w_l(t)+w_k(t)\;.
\]
If $k\neq 0$ and $l= 0$, then $w_k(t+1)=0$. 
This dynamics in words:
The mass of the main peak (located at the 
origin) is not varied. Instead, this peak randomly emits ``daughters'' of mass one. These first-order particles 
diffuse and coalesce.
They disappear when the main peak absorbs them.

This model yields positions of the 1st order peaks that crowd around the main peak but can also leave its vicinity,
making excursions to the ``bulk'' of the lattice. Note also that the model is not space-shift invariant
because we fix the main peak at the origin.

To characterize the crowding, it is important to distinguish two types of densities.
One is $\rho(k)=\av{w(\pm k,t)}$ (here because of symmetry around zero, sites $\pm k$ 
are considered equivalent). This is an average mass; peaks with higher mass (after one or several 
coalescence events) contribute more to this density. 

\begin{figure}[!htb]
\centering
\includegraphics[width=\columnwidth]{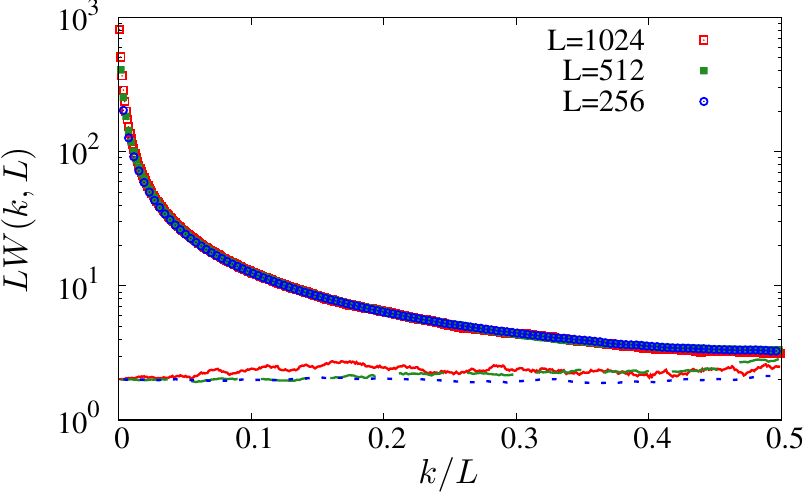}
\caption{Occupation densities $W(k,L)$ (markers) 
of the 1st order peaks along the lattice for different $L$, 
in the scaled coordinates $L\cdot W$ vs $k/L$. 
On this panel, the curves (red solid: $L=1024$; green dashed: $L=512$; blue dotted: $L=256$)
are non-normalized mass densities $\rho$ vs $k/L$.
}
\label{fig:rel-hist}
\end{figure}

\begin{figure}[ht]
\centering
\includegraphics[width=\columnwidth]{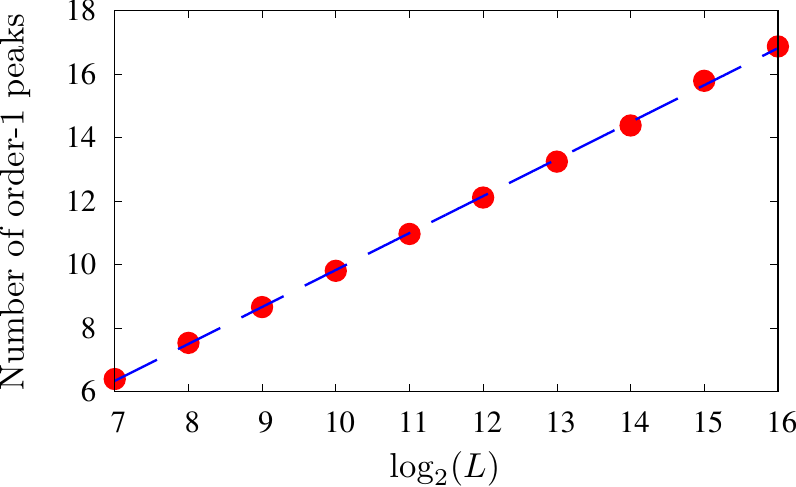}
\caption{Mean number of order-1 peaks 
as a function of lattice size $L$ (markers). The dashed line is a linear fit.}
\label{fig:1op}
\end{figure}

Another density is $W(k,t)=\av{1-\delta(w(\pm k,t))}$ where $\delta(w)=1$ if $w>0$ and $\delta(w)=1$ if $w=0$.
This density counts occupied sites, irrespective of the value of the mass $w$ at these sites.

Because the motion of the peaks between interactions with the main peak at the origin is a pure
diffusion, one expects that the stationary mass density $\rho(k)$ is uniform. On the other
hand, the occupation density $W(k)$ decays due to collisions with other 1st order peaks, thus
one expects that in the stationary state, it is maximal close to the origin and decays toward 
the bulk of the lattice.

We elaborate on statistical theory for the occupation density $W(k,t)$ in Appendix \ref{sec:apW}.
This theory predicts that the stationary density scales with the lattice size $L$ as
$W(k)=L^{-1}\hat{W}(k/L)$. This relation is tested in Fig \ref{fig:rel-hist}. We also confirm that $\rho(k)$ is uniform in this figure.

Another prediction is that the average number of the 1st-order peaks grows with
the lattice size as $\sim\log L$; this relation is checked in Fig. \ref{fig:1op}.
For higher-order peaks, the dependence on $\log L$ is nonlinear (not shown).

\section{Discussion and conclusions}
\label{sec:cscm}
We start by summarizing our main findings. We have studied the statistical properties of KR
and TT lattice models. The only two parameters are the redistribution constant $\e$ and the lattice size $L$.
There are two main regimes: $\e \lesssim L^{-1}$, where microdiffusion is small, and $\e \gtrsim L^{-1}$, where
 microdiffusion is large. For small microdiffusion, one observes a concentration of almost all mass on a single 
randomly moving site (this concentration is perfect in the KR case, $\e=0$). The masses on other sites are small, and
for very small $\e$, the dynamics builds a hierarchical structure, described in detail in Section~\ref{sec:sme}. 
For large microdiffusion, a statistically uniform in space random regime is established, as described in section
\ref{sec:lae}. In the limit $\e\to 1$ this regime corresponds to the 
Edward-Wilkinson equation with conserved noise.

It is instructive to compare the properties of the KR and TT models to other
cases from the literature where mass-conserved models, which can be interpreted as advection/transport on a one-dimensional lattice
(some with condensation/coagulation), have been studied. 

The foremost feature of all models studied in this paper and summarized
in Fig.~\ref{fig:sk} is that the average density $\rho$ is a scalable
parameter; therefore, it can't act like a control parameter and
no (equilibrium or out-of-equilibrium) phase transition can appear
when tuning $\rho$. Not even at $\rho=0$, because the model is not defined in the absence of mass. This property is, of course, shared with the random
advection-diffusion equation \eqref{eq:ade}.

Nonetheless, it is interesting to analyze our models in terms of \textit{condensation},
a process occurring when a finite fraction of the total
mass/energy is localized on a finite number of sites; in our language 
this corresponds to the formation of a maximal cluster.

Adopting a strict point of view according to which a phase transition only occurs in the thermodynamic limit,
the only model displaying condensation is KR, whose steady state corresponds to a single site
hosting the whole mass. 
However, considering the mass distribution at finite size $L$, the TT model has
remarkable features because the degree of condensation depends on
the product of the intensive parameter $\e$ and the extensive parameter $L$:
we have a sort of finite-size condensation for lattices with $L\lesssim 1/\e$. 
On this scale (which diverges for $\e\to 0$), almost all mass is concentrated on a single site (a zero-order peak),
as described in Section~\ref{sec:sme}.

We add that such finite-size condensation is attained via a dynamic coarsening,
in which the condensed fraction increases with time. This phenomenon
is visible in the roughness increase with time, as described in Sec.~\ref{sec:ttstat}.

When comparing different models, the deterministic \textit{versus} random
rule to redistribute the mass is an important feature. In the TT model, the parameter $\e$ is fixed, but it 
could be chosen randomly in the unitary interval, obtaining effectively the ``random TT model",
which has been studied (although not with this name) by Rajesh and Majumdar in Ref.~\cite{Rajesh_Majumdar_2000}
(it corresponds to their symmetric model in the limit of continuous-time dynamics). Thus we denoted it RM in 
Section~\ref{sec:dm}.
The phenomenology of such a model is fairly different from ours, which is strongly dependent on $\e L$,
a parameter that is nonsense if $\e$ is random.

\begin{figure}[!htb]
\centering
\includegraphics[width=\columnwidth]{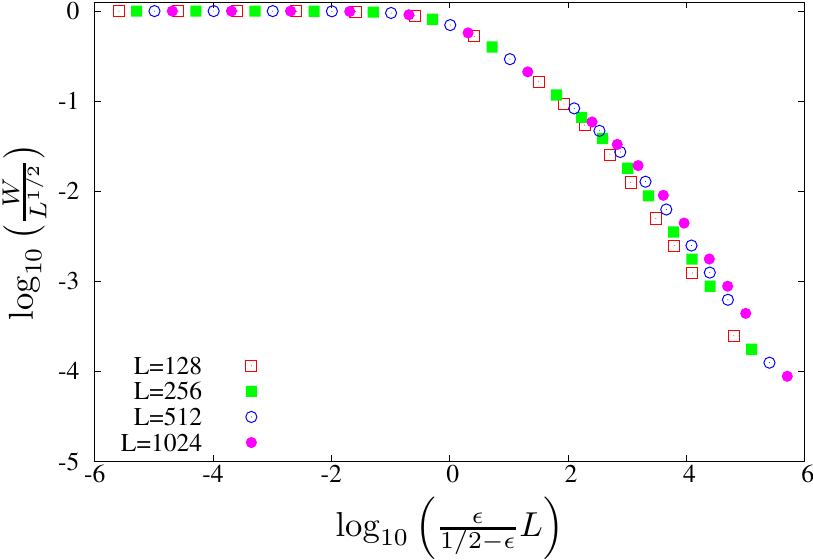}
\caption{Saturated width of the interface in the det-KMP model in scaled coordinates, 
for different $\e$ and different lattice sizes. This plot is very similar to Fig.~\ref{fig:stw}, suggesting
that the det-KMP model probably has the same statistical properties as the TT model.}
\label{fig:det-KMP}
\end{figure}

It is also instructive to compare with the
Kipnis-Marchioro-Presutti (KMP) model~\cite{kipnis1982heat}, where 
a pair of sites $(k,k\pm 1)$ is chosen randomly and the masses are redistributed according to
\begin{equation}
u'_k=\xi(u_k+u_{k\pm 1}),\quad u'_{k\pm 1}=(1-\xi)(u_k+u_{k\pm 1})\;,
\label{eq:kmp}
\end{equation}
$\xi$ being a random number uniformly distributed in $(0,1]$. 
In Ref.~\cite{kipnis1982heat}, it has been proven that a stationary distribution in the KMP model
is an equilibrium Gibbs distribution. This is not  surprising because
the update rule \eqref{eq:kmp} satisfies the detailed balance condition. This is in contradistinction
to the TT model (even in its random version), where the detailed balance is not valid. 
KMP is the prototypical example of diffusion, therefore, of a smoothing process.
It is, therefore, interesting to consider its deterministic counterpart (det-KMP in Section~\ref{sec:dm}), 
where $\xi$ is fixed, and we denote $\xi=\e$.
In the limit $\e =0$, it is the same as the KR model; we might therefore think that for small $\e$
it is similar to the TT model.
Before considering this possibility, we should note that the det-KMP model is invariant
under the transformation $\e\to 1-\e$, which sets $1/2$ as the maximal value of $\e$.
For this reason, when reproducing Fig.~\ref{fig:stw} for the det-KMP model, we have replaced
$\e/(1-\e)$ (see the label of the horizontal axis) with $\e/(0.5 -\e)$.
The result is plotted in Fig.~\ref{fig:det-KMP}, showing a strong resemblance between the two
\textit{deterministic} models.
We emphasize ``deterministic" to stress that the TT model is far more similar to det-KMP than 
to the random version of itself (RM).

It is worth stressing a further difference between the finite-size localization found in TT and det-KMP
and the standard condensation process found in models where the density $\rho$ (density of mass or other conserved
quantity) plays the role of the control parameter.
There is a plethora of such models, and it is not the case to review them, here we limit
to say that condensation exists if $\rho$ is larger than some critical value $\rho_c$
(we will give a specific example later).
Let us now try a parallel between the standard condensed phase appearing for  $\rho > \rho_c$, and the
finite-size localized phase found in TT and det-KMP models when $\e < \e_c =1/L$.
In both cases, we have a condensate in equilibrium with a background
and, for large $L$, the condensate is composed of a single site hosting a 
macroscopically large peak. 
Let us now imagine removing the condensate: in the TT model, the density is scalable, and a new condensate
will spontaneously appears. 
Instead, in the standard condensation models, this will not occur because removing the condensate will also reduce 
the density to its  critical value, $\rho \to \rho_c$.
We stress that this is not just 
a trivial reformulation due to the different physical nature of the two 
control parameters $\rho$ and $\epsilon$ (a ``density" \textit{versus} ``not a density"), 
but a consequence of a different dynamical behavior in the condensed phase.
In standard condensation models, the condensate and the background belong
to two different equilibrium phases, and the removal of the condensate does
not affect the background. In the TT model, there is really no such distinction, 
as clarified by the order kinetic model above.

To obtain a standard condensation transition where the density does play the role of a
control parameter, it is necessary to introduce a physical scale of the mass.
For this reason, we conclude this discussion by
mentioning a chipping model (CM)~\cite{Majumdar_etal-98,Rajesh_Majumdar_2001} 
where the mass $u_i$ is a non-negative \textit{integer}
and which seems to have some features similar to the TT model. 
Here mass is transported (advected) symmetrically to one of the neighboring sites with two
distinct and parallel processes:
(i)~The whole mass (i.e., all particles) is transported from site $i$ to site $i\pm 1$ (this corresponds to advection, or macrodiffusion). This occurs with rate 1.
(ii)~One single particle (if existing) is transported  from site $i$ to site $i\pm 1$ (this corresponds to microdiffusion). This occurs with rate $w$.
This model displays condensation for $\rho > \rho_c = \sqrt{1+w} - 1$.

If $w=0$, then $\rho_c =0$, and this model is equivalent to the KR model, displaying the
formation of a single cluster containing all the mass.
If $\rho\gg 1$, the discrete nature of the mass is not relevant, and for diverging $w$  
CM is similar to the TT model for $(1-\e) \ll 1$, which explains why it does not display condensation.
The CM model is another simple example of a lattice with a clear separation of micro- and macro-diffusion: 
for $w=0$ there is only macrodiffusion,
for $w=\infty$ there is only microdiffusion. For finite values of $w$, their relative importance depends on the
mass density.

In this manuscript, we have proposed a unifying picture to gather several
models of local mass transport under the same umbrella, see Fig.~\ref{fig:sk}.
It seems to us that different models, also including models
not covered by such an umbrella, e.g., the just-mentioned CM model,
might be discussed in terms of micro/macro diffusion, macrodiffusion
being a key element in obtaining a condensation-like phenomenon
and microdiffusion being the obstacle to it.
In particular, we have discussed in detail a deterministic process
where a single parameter $\e$ allows to switch between the two cases.

Since we have shown that the deterministic or random nature of the
parameters $a,b$ entering in the definition of the generic two-site model,
see Sec.~\ref{sec:dm}, is of crucial importance, it might
be of interest to distinguish between these two classes and
determine the ensemble of models of each class displaying condensation.

Another challenging problem for future studies is an extension of the models above to two- and three-dimensional lattices 
(cf.~\cite{PhysRevE.74.021124} for studies of particles sliding along two-dimensional surfaces). Here already the phenomenology of pure random advection is nontrivial, as depending on the sign of the maximal Lyapunov exponent, one can observe in absence of microdiffusion either a single cluster (delta-distribution of density like in one-dimensional case) or a random fractal. It is not clear how the latter case can be modeled on a lattice.

Finally, we mention that a possible experimental setup where statistics of one-dimensional random advection can be studied is that of particles floating on a surface of fluid where one-dimensional wave turbulence is realized~\cite{Ricard_2021,Falcon-Mordant-22}. While natural turbulence has quite specific statistical properties (that of Kolmogorov-Zakharov spectrum~\cite{zakharov2004one}), a more random field could be potentially created via external random driving. It might, however, happen that in such an experiment the cluster size will be affected not by microdiffusion, but by the finite sizes of the particles.

\acknowledgments
We thank P. Grassberger for valuable discussions. AP acknowledges hospitality
of ISC-CNR in Florence.

\appendix

\section{The TT model as an IFS}
\label{sec:bc}

  In this appendix, we demonstrate fractal properties of the invariant distribution in the TT~\eqref{eq:tt} 
  model for small lattice lengths, adopting the Iterated Function Systems (IFS) concept.
\begin{figure}[!htb]
\centering
\includegraphics[width=\columnwidth]{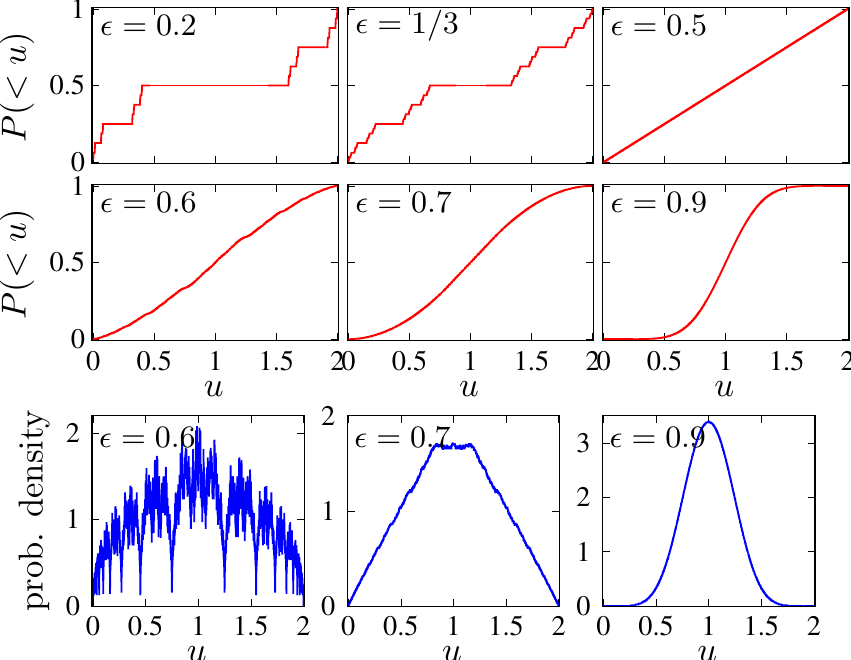}
\caption{The cumulative distribution $P(<u)$ in the IFS system \eqref{eq:ifs} 
for different $\e$ (upper two panels). This distribution is a classical fractal for $\e<1/2$.
The distribution is continuous for $\e>1/2$ but becomes smooth only for large $\e$.
Bottom panel: densities for cases $\e>1/2$.
}
\label{fig:ifsf}
\end{figure}

\subsection{Case $L=2$: Bernoulli convolution}
Let us consider the minimal case $L=2$ and $u_1+u_2=2$. Because of the conservation law, we have
just one nontrivial variable $u$, since another variable is expressed as $2-u$. In this case
the transformation \eqref{eq:upd} reads
\begin{equation}
u(t+1)=\begin{cases} \e u(t)& \text{ Prob }1/2\;,\\
\e u(t)+2(1-\e)&\text{ Prob }1/2\;.
\end{cases}
\label{eq:ifs}
\end{equation}
This one-dimensional IFS is the so-called Bernoulli convolution 
\cite{peres2000sixty,fraser1992periodic,kapral1993dynamics}.
Bernoulli convolution  generates 
a classical fractal if $\e<1/2$, and a relatively smooth distribution without voids for $\e> 1/2$, see
\cite{peres2006absolute,barral2021multifractal}. [For $\e=1/2$, the invariant distribution is uniform.]
The expression for the  dimension (there is only one dimension for $\e<1/2$, the set is a 
mono-fractal) is a trivial application of the
scaling relation: $d=-\frac{\log 2}{\log \e}$ and is smaller than one for $\e<1/2$.
In Fig.~\ref{fig:ifsf}, fractal and smooth examples are presented 
\cite{barral2021multifractal,peres2006absolute}.

It is straightforward, using linearity of \eqref{eq:ifs}, to calculate statistical properties of $u(t)$
(cf. \cite{Pikovsky-Tsimring-23}).
The average is $\av{u}=1$. For the autocorrelation function $C(t)=\av{(u(t)-1)(u(0)-1)}$ one easily obtains a 
recursion $C(t+1)=\e C(t)$, from which the exponential decay of correlations follows $C(t)=C(0)\e^t$.

\subsection{Case $L=3$}
\begin{figure}[!htb]
\centering
\includegraphics[width=\columnwidth]{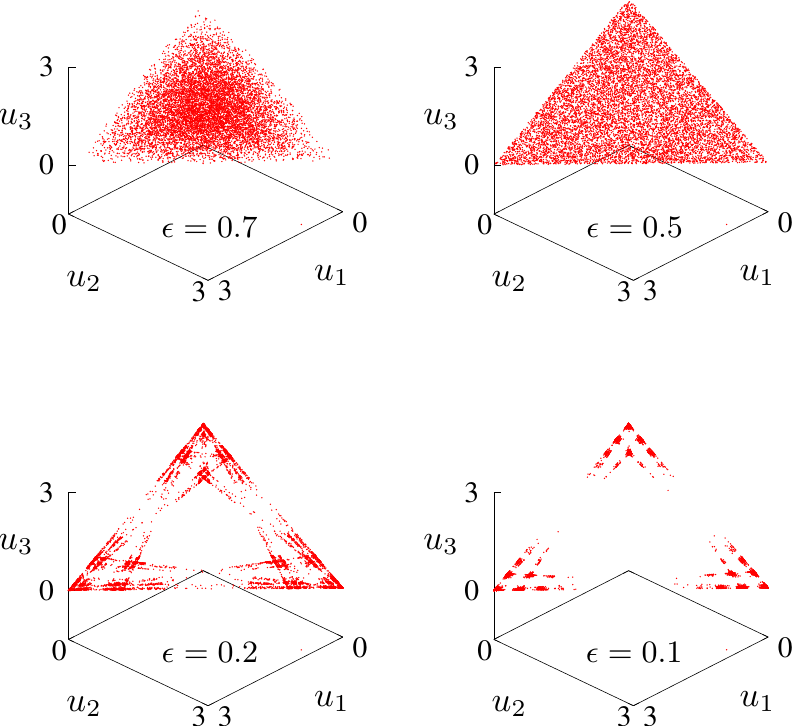}
\caption{Distributions of fields $(u_1,u_2,u_3)$ for $L=3$ and several values of $\e$.
}
\label{fig:ifs3d}
\end{figure}

In this case, because of the conservation law $u_1+u_2+u_3=3$, the dynamics 
lies on a two-dimensional simplex. Several images of the distribution ($10^4$ points are drawn)
are shown in Fig.~\ref{fig:ifs3d}. Like in the case $L=2$, the distribution is without voids for $\e\geq 0.5$
and a fractal measure with a hierarchy of voids for $\e<0.5$.

\section{Gaussianity of the field distribution in the limit $\e\to 1$}
 \label{app:gauss}
 Here we rewrite the TT model using $\mu=1-\e\ll 1$:
\[
\bar{u}=\begin{cases} (1-\mu)u& \text{ prob } 1/2\;,\\
u+\mu u_{\pm}&\text{ prob } 1/2\;.
\end{cases}
\]
Let us introduce the characteristic function $C(k)=\av{e^{iku}}$ and assume that $u$ and $u_{\pm}$ are 
statistically independent. Then the Perron-Frobenius equation for $C$ reads
\[
\bar{C}(k)=\frac{1}{2}C(k(1-\mu))+\frac{1}{2}C(k)C(k\mu)\;.
\]
In the stationary situation $\bar{C}=C$ and we obtain
\begin{equation}
2=C(\mu k)+\frac{C((1-\mu)k)}{C(k)}\;.
\label{eq:c}
\end{equation}
Because the mean value of $u$ is arbitrary, we set it to one. Then the characteristic function
can be written in terms of cumulants $\kappa_m,\;m\geq 2$:
\[
C(k)=\exp\left[ik+\sum_{m=2}^\infty\kappa_m \frac{k^m}{m!}\right]\;.
\]
For the ratio of characteristic functions, we get
\begin{gather*}
\frac{C((1-\mu)k)}{C(k)}=\exp[-ik\mu+\kappa_2\frac{k^2}{2}(-2\mu+\mu^2)+\\+
\kappa_3\frac{k^3}{6}(-3\mu+3\mu^2-\mu^3)+\ldots]\;.
\end{gather*}
The equation for $C$ \eqref{eq:c} then reads
\begin{gather*}
2=\exp[-ik\mu+\kappa_2\frac{k^2}{2}(-2\mu+\mu^2)+\\+
\kappa_3\frac{k^3}{6}(-3\mu+3\mu^2-\mu^3)+\ldots]+\\
+\exp[ik\mu+\kappa_2\mu^2\frac{k^2}{2}+\kappa_3\mu^3\frac{k^3}{6}+\ldots]\;.
\end{gather*}
We now expand the r.h.s. keeping orders $\mu^0,\mu^1,\mu^2$ only:
\begin{gather*}
2=2+\mu[-\kappa_2k^2-\kappa_3\frac{k^3}{2}-\kappa_4\frac{k^4}{6}-\ldots]-\mu^2k^2+\\
\mu^2[\kappa_2\frac{k^2}{2}+\kappa_3\frac{k^3}{2}+\kappa_4\frac{k^4}{4}+\ldots]
\end{gather*}
Comparing terms at $k^2$, in the leading order in $\mu$, we get $\kappa_2=\mu$. Comparing terms at 
$k^3,k^4,\ldots$ we get $\kappa_3=\kappa_4=\ldots=0$. This proves that for $\mu\to 0$ 
the field is Gaussian, with 
the variance $\kappa_2=\mu$.

 \section{Statistical theory for the occupation density of first-order peaks}
 \label{sec:apW}

\begin{figure}[ht]
\centering
\includegraphics[width=\columnwidth]{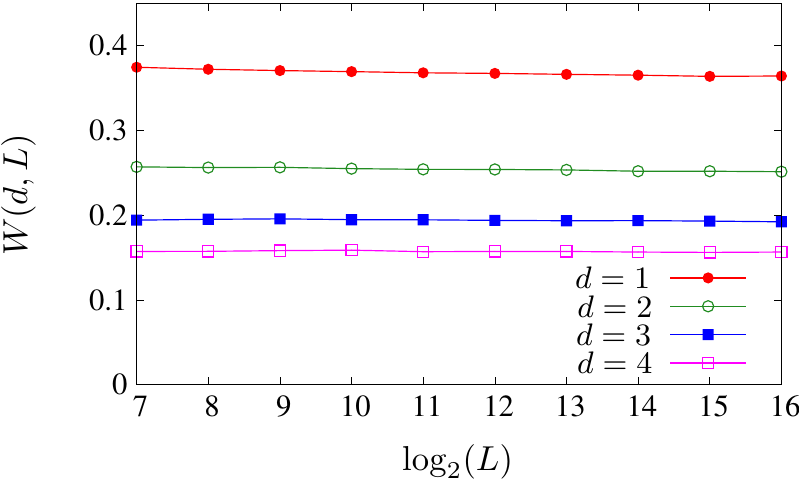}
\caption{Mean occupations of the sites $W(k=d)$ close to the main peak (which is at $k=0$), 
for different
sizes of the lattice $L$. The closest neighbor site has occupation $\approx 0.37$.}
\label{fig:hist-ad}
\end{figure}

Here we derive, within the OKM, an equation for the evolution of the occupation density of first-order peaks 
(treated as particles) $W(k,t)$,
in a lattice of length $L$.
First, we replace $k$ with a continuous coordinate $0\leq x\leq L$.
The basic dynamics is a diffusion of particles; thus, we start with the diffusion equation $\partial_t W(x,t)=D\partial_{xx}W(x,t)$.

Due to the coalescence of particles, the number of occupation sites decreases. Therefore one should add a damping term.
To elaborate on it, consider a spatially homogeneous state with constant density $W$.
Then the distance
between the particles is $X\approx W^{-1}$. The time for two particles to coalesce is the 
diffusion time for the distance between the peaks, i.e.
$t_c\approx X^2/D$. The rate of coalescence is therefore $t_c^{-1}=DX^{-2}=DW^2$. Thus from the equation
$\partial_t (\ln W)=-DW^2$
we get the damping term
$\partial_t W=-DW^3
$.
We notice that the spatially homogeneous solution
$
W(t)=\frac{W(0)}{\sqrt{1+2DW^2(0)(t-t_0)}}
$
yields the asymptotic time dependence $W\sim t^{-1/2}$, which is confirmed by numerics.

However, in our case, the occupation density is inhomogeneous because particles are
created at $x=0$ (or, equivalently, at $x=L$), where the main peak is placed.
The sites near the main peak are predominantly occupied; therefore, we can assume a constant
occupation $W=c$ at this boundary. To check that the density at the boundary does not depend on $L$, we 
calculated this density numerically at sites adjacent to  the main peak in the TT model; see results in 
Fig.~\ref{fig:hist-ad}. This figure shows that the occupation
of the nearest neighbor to the peak is around $0.37$, almost independent on $L$.

Summarizing, we have the following PDE and boundary conditions for the occupation density $W(x,t)$:
\[
\frac{\partial W(x,t)}{\partial t}=D\frac{\partial^2 W}{\partial x^2}-DW^3,\quad W(0,t)=W(L,t)=c
\]
for some $c>0$.
The stationary solution $W(x)$ of this equation obeys
\[
\frac{d^2 W}{d x^2}-W^3=0,\qquad W(0)=W(L)=c\;.
\]
We can integrate once and obtain
\begin{equation}
\begin{gathered}
\frac{1}{2}\left(\frac{dW}{dx}\right)^2-\frac{1}{4}W^4=-\frac{w^4}{4}\,;\\
\frac{dW}{dx}=\pm \frac{1}{\sqrt{2}}\sqrt{W^4-w^4}\;,
\end{gathered}
\label{eq:wx}
\end{equation}
where we took as a constant $w=W_{min}=W(L/2)$.
Integration of this equation on the interval $0\leq x \leq L/2$ yields
\[
\frac{w L}{2\sqrt{2}}=\int_1^{cw^{-1}}\frac{dy}{\sqrt{y^4-1}}=\frac{1}{\sqrt{2}}F(\arccos \frac{w}{c},\sqrt{2}/2)\;,
\]
where $F(\phi,k)$ is the elliptic integral of the 1st kind. The constant $w$ should be obtained
self-consistently from this relation. If we assume $w\ll 1$, which is to be expected for large lattice sizes $L$,
then $F(0,k)=\pi/2$
and we get a relation
\[
wL=\pi\;.
\]
Substituting this in \eqref{eq:wx}, we can represent the stationary occupation density 
as $W(x)=\frac{\pi}{L}Q(\frac{x}{L})$, where $Q(z)$ is a solution of ODE 
$\frac{dQ}{dz}=-\pi2^{-1/2}\sqrt{Q^4-1}$ with initial condition $Q(0)=cL\pi^{-1}$. 
This relation agrees with data in Fig.~\ref{fig:rel-hist}(b).

We can calculate the average number of particles in the lattice from the derived distribution.
The total number of order-1 peaks is
\begin{gather*}
K=2\int_0^{L/2} W(x) dx=2\sqrt{2}\int_w^c \frac{W dW}{\sqrt{W^4-w^4}}=\\
\sqrt{2}\int_1^{c^2 w^{-2}}\frac{dy}{\sqrt{y^2-1}}=\left.2^{1/2}\ln(y+\sqrt{y^2-1})\right|_1^{c^2w^{-2}}\\
\approx 2^{3/2}\ln L\;,
\end{gather*}
where in the last expression we assumed $w\ll 1$ and neglected $\ln c\pi$. The numerical dependence of $K$ on $\ln L$
is presented in Fig.~\ref{fig:1op}.

\bibliography{randadv1}

\begin{thebibliography}{60}%
\makeatletter
\providecommand \@ifxundefined [1]{%
 \@ifx{#1\undefined}
}%
\providecommand \@ifnum [1]{%
 \ifnum #1\expandafter \@firstoftwo
 \else \expandafter \@secondoftwo
 \fi
}%
\providecommand \@ifx [1]{%
 \ifx #1\expandafter \@firstoftwo
 \else \expandafter \@secondoftwo
 \fi
}%
\providecommand \natexlab [1]{#1}%
\providecommand \enquote  [1]{``#1''}%
\providecommand \bibnamefont  [1]{#1}%
\providecommand \bibfnamefont [1]{#1}%
\providecommand \citenamefont [1]{#1}%
\providecommand \href@noop [0]{\@secondoftwo}%
\providecommand \href [0]{\begingroup \@sanitize@url \@href}%
\providecommand \@href[1]{\@@startlink{#1}\@@href}%
\providecommand \@@href[1]{\endgroup#1\@@endlink}%
\providecommand \@sanitize@url [0]{\catcode `\\12\catcode `\$12\catcode
  `\&12\catcode `\#12\catcode `\^12\catcode `\_12\catcode `\%12\relax}%
\providecommand \@@startlink[1]{}%
\providecommand \@@endlink[0]{}%
\providecommand \url  [0]{\begingroup\@sanitize@url \@url }%
\providecommand \@url [1]{\endgroup\@href {#1}{\urlprefix }}%
\providecommand \urlprefix  [0]{URL }%
\providecommand \Eprint [0]{\href }%
\providecommand \doibase [0]{https://doi.org/}%
\providecommand \selectlanguage [0]{\@gobble}%
\providecommand \bibinfo  [0]{\@secondoftwo}%
\providecommand \bibfield  [0]{\@secondoftwo}%
\providecommand \translation [1]{[#1]}%
\providecommand \BibitemOpen [0]{}%
\providecommand \bibitemStop [0]{}%
\providecommand \bibitemNoStop [0]{.\EOS\space}%
\providecommand \EOS [0]{\spacefactor3000\relax}%
\providecommand \BibitemShut  [1]{\csname bibitem#1\endcsname}%
\let\auto@bib@innerbib\@empty
\bibitem [{\citenamefont {Warhaft}(2000)}]{warhaft2000passive}%
  \BibitemOpen
  \bibfield  {author} {\bibinfo {author} {\bibfnamefont {Z.}~\bibnamefont
  {Warhaft}},\ }\bibfield  {title} {\bibinfo {title} {Passive scalars in
  turbulent flows},\ }\href@noop {} {\bibfield  {journal} {\bibinfo  {journal}
  {Annual Review of Fluid Mechanics}\ }\textbf {\bibinfo {volume} {32}},\
  \bibinfo {pages} {203} (\bibinfo {year} {2000})}\BibitemShut {NoStop}%
\bibitem [{\citenamefont {Kraichnan}(1994)}]{kraichnan1994anomalous}%
  \BibitemOpen
  \bibfield  {author} {\bibinfo {author} {\bibfnamefont {R.~H.}\ \bibnamefont
  {Kraichnan}},\ }\bibfield  {title} {\bibinfo {title} {Anomalous scaling of a
  randomly advected passive scalar},\ }\href
  {https://doi.org/10.1103/PhysRevLett.72.1016} {\bibfield  {journal} {\bibinfo
   {journal} {Phys. Rev. Lett.}\ }\textbf {\bibinfo {volume} {72}},\ \bibinfo
  {pages} {1016} (\bibinfo {year} {1994})}\BibitemShut {NoStop}%
\bibitem [{\citenamefont {Elperin}\ \emph {et~al.}(1995)\citenamefont
  {Elperin}, \citenamefont {Kleeorin},\ and\ \citenamefont
  {Rogachevskii}}]{Elperin_etal-95}%
  \BibitemOpen
  \bibfield  {author} {\bibinfo {author} {\bibfnamefont {T.}~\bibnamefont
  {Elperin}}, \bibinfo {author} {\bibfnamefont {N.}~\bibnamefont {Kleeorin}},\
  and\ \bibinfo {author} {\bibfnamefont {I.}~\bibnamefont {Rogachevskii}},\
  }\bibfield  {title} {\bibinfo {title} {Dynamics of the passive scalar in
  compressible turbulent flow: {L}arge-scale patterns and small-scale
  fluctuations},\ }\href {https://doi.org/10.1103/PhysRevE.52.2617} {\bibfield
  {journal} {\bibinfo  {journal} {Phys. Rev. E}\ }\textbf {\bibinfo {volume}
  {52}},\ \bibinfo {pages} {2617} (\bibinfo {year} {1995})}\BibitemShut
  {NoStop}%
\bibitem [{\citenamefont {Vulpiani}\ \emph {et~al.}(2009)\citenamefont
  {Vulpiani}, \citenamefont {Cecconi},\ and\ \citenamefont
  {Cencini}}]{vulpiani2009chaos}%
  \BibitemOpen
  \bibfield  {author} {\bibinfo {author} {\bibfnamefont {A.}~\bibnamefont
  {Vulpiani}}, \bibinfo {author} {\bibfnamefont {F.}~\bibnamefont {Cecconi}},\
  and\ \bibinfo {author} {\bibfnamefont {M.}~\bibnamefont {Cencini}},\
  }\href@noop {} {\emph {\bibinfo {title} {Chaos: from simple models to complex
  systems}}},\ Vol.~\bibinfo {volume} {17}\ (\bibinfo  {publisher} {World
  Scientific},\ \bibinfo {year} {2009})\BibitemShut {NoStop}%
\bibitem [{\citenamefont {Deutsch}(1994)}]{deutsch1994probability}%
  \BibitemOpen
  \bibfield  {author} {\bibinfo {author} {\bibfnamefont {J.~M.}\ \bibnamefont
  {Deutsch}},\ }\bibfield  {title} {\bibinfo {title} {Probability distributions
  for multicomponent systems with multiplicative noise},\ }\href@noop {}
  {\bibfield  {journal} {\bibinfo  {journal} {Physica A}\ }\textbf {\bibinfo
  {volume} {208}},\ \bibinfo {pages} {445} (\bibinfo {year}
  {1994})}\BibitemShut {NoStop}%
\bibitem [{\citenamefont {Lepri}(2013)}]{lepri2013fluctuations}%
  \BibitemOpen
  \bibfield  {author} {\bibinfo {author} {\bibfnamefont {S.}~\bibnamefont
  {Lepri}},\ }\bibfield  {title} {\bibinfo {title} {Fluctuations in a diffusive
  medium with gain},\ }\href@noop {} {\bibfield  {journal} {\bibinfo  {journal}
  {Phys. Rev. Lett.}\ }\textbf {\bibinfo {volume} {110}},\ \bibinfo {pages}
  {230603} (\bibinfo {year} {2013})}\BibitemShut {NoStop}%
\bibitem [{\citenamefont {Schr\"oder}\ \emph {et~al.}(1996)\citenamefont
  {Schr\"oder}, \citenamefont {Andersen}, \citenamefont {Levinsen},
  \citenamefont {Alstr\o{}m},\ and\ \citenamefont
  {Goldburg}}]{PhysRevLett.76.4717}%
  \BibitemOpen
  \bibfield  {author} {\bibinfo {author} {\bibfnamefont {E.}~\bibnamefont
  {Schr\"oder}}, \bibinfo {author} {\bibfnamefont {J.~S.}\ \bibnamefont
  {Andersen}}, \bibinfo {author} {\bibfnamefont {M.~T.}\ \bibnamefont
  {Levinsen}}, \bibinfo {author} {\bibfnamefont {P.}~\bibnamefont
  {Alstr\o{}m}},\ and\ \bibinfo {author} {\bibfnamefont {W.~I.}\ \bibnamefont
  {Goldburg}},\ }\bibfield  {title} {\bibinfo {title} {Relative particle motion
  in capillary waves},\ }\href {https://doi.org/10.1103/PhysRevLett.76.4717}
  {\bibfield  {journal} {\bibinfo  {journal} {Phys. Rev. Lett.}\ }\textbf
  {\bibinfo {volume} {76}},\ \bibinfo {pages} {4717} (\bibinfo {year}
  {1996})}\BibitemShut {NoStop}%
\bibitem [{\citenamefont {Ricard}\ and\ \citenamefont
  {Falcon}(2021)}]{Ricard_2021}%
  \BibitemOpen
  \bibfield  {author} {\bibinfo {author} {\bibfnamefont {G.}~\bibnamefont
  {Ricard}}\ and\ \bibinfo {author} {\bibfnamefont {E.}~\bibnamefont
  {Falcon}},\ }\bibfield  {title} {\bibinfo {title} {Experimental quasi-1d
  capillary-wave turbulence},\ }\href
  {https://doi.org/10.1209/0295-5075/ac2751} {\bibfield  {journal} {\bibinfo
  {journal} {Europhysics Letters}\ }\textbf {\bibinfo {volume} {135}},\
  \bibinfo {pages} {64001} (\bibinfo {year} {2021})}\BibitemShut {NoStop}%
\bibitem [{\citenamefont {Leyvraz}(2003)}]{leyvraz2003scaling}%
  \BibitemOpen
  \bibfield  {author} {\bibinfo {author} {\bibfnamefont {F.}~\bibnamefont
  {Leyvraz}},\ }\bibfield  {title} {\bibinfo {title} {Scaling theory and
  exactly solved models in the kinetics of irreversible aggregation},\ }\href
  {https://doi.org/https://doi.org/10.1016/S0370-1573(03)00241-2} {\bibfield
  {journal} {\bibinfo  {journal} {Physics Reports}\ }\textbf {\bibinfo {volume}
  {383}},\ \bibinfo {pages} {95} (\bibinfo {year} {2003})}\BibitemShut
  {NoStop}%
\bibitem [{\citenamefont {Pikovsky}(1984)}]{Pikovsky-84a}%
  \BibitemOpen
  \bibfield  {author} {\bibinfo {author} {\bibfnamefont {A.~S.}\ \bibnamefont
  {Pikovsky}},\ }\bibfield  {title} {\bibinfo {title} {Synchronization and
  stochastization of the ensemble of autogenerators by external noise},\
  }\href@noop {} {\bibfield  {journal} {\bibinfo  {journal} {Radiophys. Quantum
  Electron.}\ }\textbf {\bibinfo {volume} {27}},\ \bibinfo {pages} {390}
  (\bibinfo {year} {1984})}\BibitemShut {NoStop}%
\bibitem [{\citenamefont {Antonov}(1984)}]{antonov1984}%
  \BibitemOpen
  \bibfield  {author} {\bibinfo {author} {\bibfnamefont {V.~A.}\ \bibnamefont
  {Antonov}},\ }\bibfield  {title} {\bibinfo {title} {Modeling of processes of
  cyclic evolution type. synchronization by a random signal.},\ }\href@noop {}
  {\bibfield  {journal} {\bibinfo  {journal} {Proceedings of Leningrad
  University, Astronomy (in Russian)}\ ,\ \bibinfo {pages} {67}} (\bibinfo
  {year} {1984})}\BibitemShut {NoStop}%
\bibitem [{\citenamefont {Crauel}\ and\ \citenamefont
  {Flandoli}(1994)}]{Crauel-Flandoli-94}%
  \BibitemOpen
  \bibfield  {author} {\bibinfo {author} {\bibfnamefont {H.}~\bibnamefont
  {Crauel}}\ and\ \bibinfo {author} {\bibfnamefont {F.}~\bibnamefont
  {Flandoli}},\ }\bibfield  {title} {\bibinfo {title} {Attractors for random
  dynamical systems},\ }\href@noop {} {\bibfield  {journal} {\bibinfo
  {journal} {Probab. Theory Relat. Fields}\ }\textbf {\bibinfo {volume}
  {100}},\ \bibinfo {pages} {365} (\bibinfo {year} {1994})}\BibitemShut
  {NoStop}%
\bibitem [{\citenamefont {Mainen}\ and\ \citenamefont
  {Sejnowski}(1995)}]{Mainen-Sejnowski-95}%
  \BibitemOpen
  \bibfield  {author} {\bibinfo {author} {\bibfnamefont {Z.~F.}\ \bibnamefont
  {Mainen}}\ and\ \bibinfo {author} {\bibfnamefont {T.~J.}\ \bibnamefont
  {Sejnowski}},\ }\bibfield  {title} {\bibinfo {title} {Reliability of spike
  timing in neocortical neurons},\ }\href@noop {} {\bibfield  {journal}
  {\bibinfo  {journal} {Science}\ }\textbf {\bibinfo {volume} {268}},\ \bibinfo
  {pages} {1503} (\bibinfo {year} {1995})}\BibitemShut {NoStop}%
\bibitem [{\citenamefont {Khoury}\ \emph {et~al.}(1998)\citenamefont {Khoury},
  \citenamefont {Lieberman},\ and\ \citenamefont {Lichtenberg}}]{Khoury-98}%
  \BibitemOpen
  \bibfield  {author} {\bibinfo {author} {\bibfnamefont {P.}~\bibnamefont
  {Khoury}}, \bibinfo {author} {\bibfnamefont {M.~A.}\ \bibnamefont
  {Lieberman}},\ and\ \bibinfo {author} {\bibfnamefont {A.~J.}\ \bibnamefont
  {Lichtenberg}},\ }\bibfield  {title} {\bibinfo {title} {Experimental
  measurement of the degree of chaotic synchronization using a distribution
  exponent},\ }\href@noop {} {\bibfield  {journal} {\bibinfo  {journal} {Phys.
  Rev. E}\ }\textbf {\bibinfo {volume} {57}},\ \bibinfo {pages} {5448}
  (\bibinfo {year} {1998})}\BibitemShut {NoStop}%
\bibitem [{\citenamefont {Uchida}\ \emph {et~al.}(2004)\citenamefont {Uchida},
  \citenamefont {McAllister},\ and\ \citenamefont
  {Roy}}]{Uchida-Mcallister-Roy-04}%
  \BibitemOpen
  \bibfield  {author} {\bibinfo {author} {\bibfnamefont {A.}~\bibnamefont
  {Uchida}}, \bibinfo {author} {\bibfnamefont {R.}~\bibnamefont {McAllister}},\
  and\ \bibinfo {author} {\bibfnamefont {R.}~\bibnamefont {Roy}},\ }\bibfield
  {title} {\bibinfo {title} {Consistency of nonlinear system response to
  complex drive signals},\ }\href@noop {} {\bibfield  {journal} {\bibinfo
  {journal} {Phys. Rev. Lett.}\ }\textbf {\bibinfo {volume} {93}},\ \bibinfo
  {pages} {244102} (\bibinfo {year} {2004})}\BibitemShut {NoStop}%
\bibitem [{\citenamefont {Pikovsky}(1992)}]{Pikovsky-92c}%
  \BibitemOpen
  \bibfield  {author} {\bibinfo {author} {\bibfnamefont {A.~S.}\ \bibnamefont
  {Pikovsky}},\ }\bibfield  {title} {\bibinfo {title} {Statistics of trajectory
  separation in noisy dynamical systems},\ }\href@noop {} {\bibfield  {journal}
  {\bibinfo  {journal} {Phys. Lett. A}\ }\textbf {\bibinfo {volume} {165}},\
  \bibinfo {pages} {33} (\bibinfo {year} {1992})}\BibitemShut {NoStop}%
\bibitem [{\citenamefont {Sommerer}\ and\ \citenamefont
  {Ott}(1993)}]{Sommerer-Ott-93a}%
  \BibitemOpen
  \bibfield  {author} {\bibinfo {author} {\bibfnamefont {J.~C.}\ \bibnamefont
  {Sommerer}}\ and\ \bibinfo {author} {\bibfnamefont {E.}~\bibnamefont {Ott}},\
  }\bibfield  {title} {\bibinfo {title} {Particles floating on a moving fluid:
  a dynamically comprehensible physical fractal},\ }\href@noop {} {\bibfield
  {journal} {\bibinfo  {journal} {Science}\ }\textbf {\bibinfo {volume}
  {259}},\ \bibinfo {pages} {335} (\bibinfo {year} {1993})}\BibitemShut
  {NoStop}%
\bibitem [{\citenamefont {Gawedzki}\ and\ \citenamefont
  {Vergassola}(2000)}]{Gawedzki-Vergassola-00}%
  \BibitemOpen
  \bibfield  {author} {\bibinfo {author} {\bibfnamefont {K.}~\bibnamefont
  {Gawedzki}}\ and\ \bibinfo {author} {\bibfnamefont {M.}~\bibnamefont
  {Vergassola}},\ }\bibfield  {title} {\bibinfo {title} {Phase transition in
  the passive scalar advection},\ }\href@noop {} {\bibfield  {journal}
  {\bibinfo  {journal} {Physica D}\ }\textbf {\bibinfo {volume} {138}},\
  \bibinfo {pages} {63} (\bibinfo {year} {2000})}\BibitemShut {NoStop}%
\bibitem [{\citenamefont {Bohr}\ and\ \citenamefont
  {Pikovsky}(1993)}]{Bohr-Pikovsky-93}%
  \BibitemOpen
  \bibfield  {author} {\bibinfo {author} {\bibfnamefont {T.}~\bibnamefont
  {Bohr}}\ and\ \bibinfo {author} {\bibfnamefont {A.~S.}\ \bibnamefont
  {Pikovsky}},\ }\bibfield  {title} {\bibinfo {title} {Anomalous diffusion in
  the {K}uramoto -- {S}ivashinsky equation},\ }\href@noop {} {\bibfield
  {journal} {\bibinfo  {journal} {Phys. Rev. Lett.}\ }\textbf {\bibinfo
  {volume} {70}},\ \bibinfo {pages} {2892} (\bibinfo {year}
  {1993})}\BibitemShut {NoStop}%
\bibitem [{\citenamefont {Wang}\ and\ \citenamefont
  {Wang}(1994)}]{PhysRevE.49.5853}%
  \BibitemOpen
  \bibfield  {author} {\bibinfo {author} {\bibfnamefont {X.-H.}\ \bibnamefont
  {Wang}}\ and\ \bibinfo {author} {\bibfnamefont {K.-L.}\ \bibnamefont
  {Wang}},\ }\bibfield  {title} {\bibinfo {title} {Analysis of anomalous
  diffusion in the {K}uramoto-{S}ivashinsky equation},\ }\href
  {https://doi.org/10.1103/PhysRevE.49.5853} {\bibfield  {journal} {\bibinfo
  {journal} {Phys. Rev. E}\ }\textbf {\bibinfo {volume} {49}},\ \bibinfo
  {pages} {5853} (\bibinfo {year} {1994})}\BibitemShut {NoStop}%
\bibitem [{\citenamefont {Edwards}\ and\ \citenamefont
  {Wilkinson}(1982)}]{Edwards-Wilkinson-82}%
  \BibitemOpen
  \bibfield  {author} {\bibinfo {author} {\bibfnamefont {S.~F.}\ \bibnamefont
  {Edwards}}\ and\ \bibinfo {author} {\bibfnamefont {D.~R.}\ \bibnamefont
  {Wilkinson}},\ }\bibfield  {title} {\bibinfo {title} {The surface statistics
  of a granular aggregate},\ }\href@noop {} {\bibfield  {journal} {\bibinfo
  {journal} {Proc. Royal Soc. London}\ }\textbf {\bibinfo {volume} {A381}},\
  \bibinfo {pages} {17} (\bibinfo {year} {1982})}\BibitemShut {NoStop}%
\bibitem [{\citenamefont {Kardar}\ \emph {et~al.}(1986)\citenamefont {Kardar},
  \citenamefont {Parisi},\ and\ \citenamefont {Zhang}}]{PhysRevLett.56.889}%
  \BibitemOpen
  \bibfield  {author} {\bibinfo {author} {\bibfnamefont {M.}~\bibnamefont
  {Kardar}}, \bibinfo {author} {\bibfnamefont {G.}~\bibnamefont {Parisi}},\
  and\ \bibinfo {author} {\bibfnamefont {Y.-C.}\ \bibnamefont {Zhang}},\
  }\bibfield  {title} {\bibinfo {title} {Dynamic scaling of growing
  interfaces},\ }\href {https://doi.org/10.1103/PhysRevLett.56.889} {\bibfield
  {journal} {\bibinfo  {journal} {Phys. Rev. Lett.}\ }\textbf {\bibinfo
  {volume} {56}},\ \bibinfo {pages} {889} (\bibinfo {year} {1986})}\BibitemShut
  {NoStop}%
\bibitem [{\citenamefont {Das}\ and\ \citenamefont
  {Barma}(2000)}]{PhysRevLett.85.1602}%
  \BibitemOpen
  \bibfield  {author} {\bibinfo {author} {\bibfnamefont {D.}~\bibnamefont
  {Das}}\ and\ \bibinfo {author} {\bibfnamefont {M.}~\bibnamefont {Barma}},\
  }\bibfield  {title} {\bibinfo {title} {Particles sliding on a fluctuating
  surface: Phase separation and power laws},\ }\href
  {https://doi.org/10.1103/PhysRevLett.85.1602} {\bibfield  {journal} {\bibinfo
   {journal} {Phys. Rev. Lett.}\ }\textbf {\bibinfo {volume} {85}},\ \bibinfo
  {pages} {1602} (\bibinfo {year} {2000})}\BibitemShut {NoStop}%
\bibitem [{\citenamefont {Drossel}\ and\ \citenamefont
  {Kardar}(2002)}]{PhysRevB.66.195414}%
  \BibitemOpen
  \bibfield  {author} {\bibinfo {author} {\bibfnamefont {B.}~\bibnamefont
  {Drossel}}\ and\ \bibinfo {author} {\bibfnamefont {M.}~\bibnamefont
  {Kardar}},\ }\bibfield  {title} {\bibinfo {title} {Passive sliders on growing
  surfaces and advection in {B}urger's flows},\ }\href
  {https://doi.org/10.1103/PhysRevB.66.195414} {\bibfield  {journal} {\bibinfo
  {journal} {Phys. Rev. B}\ }\textbf {\bibinfo {volume} {66}},\ \bibinfo
  {pages} {195414} (\bibinfo {year} {2002})}\BibitemShut {NoStop}%
\bibitem [{\citenamefont {Das}\ \emph {et~al.}(2001)\citenamefont {Das},
  \citenamefont {Barma},\ and\ \citenamefont {Majumdar}}]{PhysRevE.64.046126}%
  \BibitemOpen
  \bibfield  {author} {\bibinfo {author} {\bibfnamefont {D.}~\bibnamefont
  {Das}}, \bibinfo {author} {\bibfnamefont {M.}~\bibnamefont {Barma}},\ and\
  \bibinfo {author} {\bibfnamefont {S.~N.}\ \bibnamefont {Majumdar}},\
  }\bibfield  {title} {\bibinfo {title} {Fluctuation-dominated phase ordering
  driven by stochastically evolving surfaces: {D}epth models and sliding
  particles},\ }\href {https://doi.org/10.1103/PhysRevE.64.046126} {\bibfield
  {journal} {\bibinfo  {journal} {Phys. Rev. E}\ }\textbf {\bibinfo {volume}
  {64}},\ \bibinfo {pages} {046126} (\bibinfo {year} {2001})}\BibitemShut
  {NoStop}%
\bibitem [{\citenamefont {Chin}(2002)}]{PhysRevE.66.021104}%
  \BibitemOpen
  \bibfield  {author} {\bibinfo {author} {\bibfnamefont {C.-S.}\ \bibnamefont
  {Chin}},\ }\bibfield  {title} {\bibinfo {title} {Passive random walkers and
  riverlike networks on growing surfaces},\ }\href
  {https://doi.org/10.1103/PhysRevE.66.021104} {\bibfield  {journal} {\bibinfo
  {journal} {Phys. Rev. E}\ }\textbf {\bibinfo {volume} {66}},\ \bibinfo
  {pages} {021104} (\bibinfo {year} {2002})}\BibitemShut {NoStop}%
\bibitem [{\citenamefont {Nagar}\ \emph {et~al.}(2006)\citenamefont {Nagar},
  \citenamefont {Majumdar},\ and\ \citenamefont {Barma}}]{PhysRevE.74.021124}%
  \BibitemOpen
  \bibfield  {author} {\bibinfo {author} {\bibfnamefont {A.}~\bibnamefont
  {Nagar}}, \bibinfo {author} {\bibfnamefont {S.~N.}\ \bibnamefont
  {Majumdar}},\ and\ \bibinfo {author} {\bibfnamefont {M.}~\bibnamefont
  {Barma}},\ }\bibfield  {title} {\bibinfo {title} {Strong clustering of
  noninteracting, sliding passive scalars driven by fluctuating surfaces},\
  }\href {https://doi.org/10.1103/PhysRevE.74.021124} {\bibfield  {journal}
  {\bibinfo  {journal} {Phys. Rev. E}\ }\textbf {\bibinfo {volume} {74}},\
  \bibinfo {pages} {021124} (\bibinfo {year} {2006})}\BibitemShut {NoStop}%
\bibitem [{\citenamefont {Singha}\ and\ \citenamefont
  {Barma}(2018)}]{PhysRevE.98.052148}%
  \BibitemOpen
  \bibfield  {author} {\bibinfo {author} {\bibfnamefont {T.}~\bibnamefont
  {Singha}}\ and\ \bibinfo {author} {\bibfnamefont {M.}~\bibnamefont {Barma}},\
  }\bibfield  {title} {\bibinfo {title} {Clustering, intermittency, and scaling
  for passive particles on fluctuating surfaces},\ }\href
  {https://doi.org/10.1103/PhysRevE.98.052148} {\bibfield  {journal} {\bibinfo
  {journal} {Phys. Rev. E}\ }\textbf {\bibinfo {volume} {98}},\ \bibinfo
  {pages} {052148} (\bibinfo {year} {2018})}\BibitemShut {NoStop}%
\bibitem [{\citenamefont {Zakharov}\ \emph {et~al.}(2004)\citenamefont
  {Zakharov}, \citenamefont {Dias},\ and\ \citenamefont
  {Pushkarev}}]{zakharov2004one}%
  \BibitemOpen
  \bibfield  {author} {\bibinfo {author} {\bibfnamefont {V.}~\bibnamefont
  {Zakharov}}, \bibinfo {author} {\bibfnamefont {F.}~\bibnamefont {Dias}},\
  and\ \bibinfo {author} {\bibfnamefont {A.}~\bibnamefont {Pushkarev}},\
  }\bibfield  {title} {\bibinfo {title} {One-dimensional wave turbulence},\
  }\href@noop {} {\bibfield  {journal} {\bibinfo  {journal} {Physics Reports}\
  }\textbf {\bibinfo {volume} {398}},\ \bibinfo {pages} {1} (\bibinfo {year}
  {2004})}\BibitemShut {NoStop}%
\bibitem [{\citenamefont {Majda}\ \emph {et~al.}(1997)\citenamefont {Majda},
  \citenamefont {McLaughlin},\ and\ \citenamefont {Tabak}}]{MMT}%
  \BibitemOpen
  \bibfield  {author} {\bibinfo {author} {\bibfnamefont {A.}~\bibnamefont
  {Majda}}, \bibinfo {author} {\bibfnamefont {D.}~\bibnamefont {McLaughlin}},\
  and\ \bibinfo {author} {\bibfnamefont {E.}~\bibnamefont {Tabak}},\ }\bibfield
   {title} {\bibinfo {title} {A one-dimensional model for dispersive wave
  turbulence},\ }\href@noop {} {\bibfield  {journal} {\bibinfo  {journal} {J.
  Nonlinear Sci.}\ }\textbf {\bibinfo {volume} {7}},\ \bibinfo {pages} {9}
  (\bibinfo {year} {1997})}\BibitemShut {NoStop}%
\bibitem [{\citenamefont {Cagnetta}\ \emph {et~al.}(2019)\citenamefont
  {Cagnetta}, \citenamefont {Evans},\ and\ \citenamefont
  {Marenduzzo}}]{PhysRevE.99.042124}%
  \BibitemOpen
  \bibfield  {author} {\bibinfo {author} {\bibfnamefont {F.}~\bibnamefont
  {Cagnetta}}, \bibinfo {author} {\bibfnamefont {M.~R.}\ \bibnamefont
  {Evans}},\ and\ \bibinfo {author} {\bibfnamefont {D.}~\bibnamefont
  {Marenduzzo}},\ }\bibfield  {title} {\bibinfo {title} {Statistical mechanics
  of a single active slider on a fluctuating interface},\ }\href
  {https://doi.org/10.1103/PhysRevE.99.042124} {\bibfield  {journal} {\bibinfo
  {journal} {Phys. Rev. E}\ }\textbf {\bibinfo {volume} {99}},\ \bibinfo
  {pages} {042124} (\bibinfo {year} {2019})}\BibitemShut {NoStop}%
\bibitem [{\citenamefont {Kang}\ and\ \citenamefont
  {Redner}(1984)}]{kang1984fluctuation}%
  \BibitemOpen
  \bibfield  {author} {\bibinfo {author} {\bibfnamefont {K.}~\bibnamefont
  {Kang}}\ and\ \bibinfo {author} {\bibfnamefont {S.}~\bibnamefont {Redner}},\
  }\bibfield  {title} {\bibinfo {title} {Fluctuation effects in {Smoluchowski}
  reaction kinetics},\ }\href@noop {} {\bibfield  {journal} {\bibinfo
  {journal} {Phys. Rev. A}\ }\textbf {\bibinfo {volume} {30}},\ \bibinfo
  {pages} {2833} (\bibinfo {year} {1984})}\BibitemShut {NoStop}%
\bibitem [{\citenamefont {Takayasu}\ and\ \citenamefont
  {Taguchi}(1993)}]{takayasu1993non}%
  \BibitemOpen
  \bibfield  {author} {\bibinfo {author} {\bibfnamefont {H.}~\bibnamefont
  {Takayasu}}\ and\ \bibinfo {author} {\bibfnamefont {Y.}~\bibnamefont
  {Taguchi}},\ }\bibfield  {title} {\bibinfo {title} {Non-{Gaussian}
  distribution in random advection dynamics},\ }\href@noop {} {\bibfield
  {journal} {\bibinfo  {journal} {Phys. Rev. Lett.}\ }\textbf {\bibinfo
  {volume} {70}},\ \bibinfo {pages} {782} (\bibinfo {year} {1993})}\BibitemShut
  {NoStop}%
\bibitem [{\citenamefont {Kipnis}\ \emph {et~al.}(1982)\citenamefont {Kipnis},
  \citenamefont {Marchioro},\ and\ \citenamefont {Presutti}}]{kipnis1982heat}%
  \BibitemOpen
  \bibfield  {author} {\bibinfo {author} {\bibfnamefont {C.}~\bibnamefont
  {Kipnis}}, \bibinfo {author} {\bibfnamefont {C.}~\bibnamefont {Marchioro}},\
  and\ \bibinfo {author} {\bibfnamefont {E.}~\bibnamefont {Presutti}},\
  }\bibfield  {title} {\bibinfo {title} {Heat flow in an exactly solvable
  model},\ }\href@noop {} {\bibfield  {journal} {\bibinfo  {journal} {J. Stat.
  Phys.}\ }\textbf {\bibinfo {volume} {27}},\ \bibinfo {pages} {65} (\bibinfo
  {year} {1982})}\BibitemShut {NoStop}%
\bibitem [{\citenamefont {Rajesh}\ and\ \citenamefont
  {Majumdar}(2000)}]{Rajesh_Majumdar_2000}%
  \BibitemOpen
  \bibfield  {author} {\bibinfo {author} {\bibfnamefont {R.}~\bibnamefont
  {Rajesh}}\ and\ \bibinfo {author} {\bibfnamefont {S.~N.}\ \bibnamefont
  {Majumdar}},\ }\bibfield  {title} {\bibinfo {title} {Conserved mass models
  and particle systems in one dimension},\ }\href@noop {} {\bibfield  {journal}
  {\bibinfo  {journal} {J. Stat. Phys.}\ }\textbf {\bibinfo {volume} {99}},\
  \bibinfo {pages} {943} (\bibinfo {year} {2000})}\BibitemShut {NoStop}%
\bibitem [{\citenamefont {Krishnamurthy}\ \emph {et~al.}(2003)\citenamefont
  {Krishnamurthy}, \citenamefont {Rajesh},\ and\ \citenamefont
  {Zaboronski}}]{krishnamurthy2003persistence}%
  \BibitemOpen
  \bibfield  {author} {\bibinfo {author} {\bibfnamefont {S.}~\bibnamefont
  {Krishnamurthy}}, \bibinfo {author} {\bibfnamefont {R.}~\bibnamefont
  {Rajesh}},\ and\ \bibinfo {author} {\bibfnamefont {O.}~\bibnamefont
  {Zaboronski}},\ }\bibfield  {title} {\bibinfo {title} {Persistence properties
  of a system of coagulating and annihilating random walkers},\ }\href@noop {}
  {\bibfield  {journal} {\bibinfo  {journal} {Phys. Rev. E}\ }\textbf {\bibinfo
  {volume} {68}},\ \bibinfo {pages} {046103} (\bibinfo {year}
  {2003})}\BibitemShut {NoStop}%
\bibitem [{Note1()}]{Note1}%
  \BibitemOpen
  \bibinfo {note} {Our notation is different from the notation of the original
  paper \cite {takayasu1993non} that employed the parameter $j=1-\epsilon $.
  Our choice is due to our interest towards the limit $\epsilon \to
  0$.}\BibitemShut {Stop}%
\bibitem [{Note2()}]{Note2}%
  \BibitemOpen
  \bibinfo {note} {At first glance, this statement is not obvious because the
  limit $\epsilon \to 1$ might not commute with the thermodynamic limit $L\to
  \infty $. However, we will provide a consistent description for any $L$ and
  $\epsilon $.}\BibitemShut {Stop}%
\bibitem [{\citenamefont {Takayasu}\ \emph {et~al.}(1994)\citenamefont
  {Takayasu}, \citenamefont {Takayasu},\ and\ \citenamefont
  {Taguchi}}]{takayasu1994non}%
  \BibitemOpen
  \bibfield  {author} {\bibinfo {author} {\bibfnamefont {M.}~\bibnamefont
  {Takayasu}}, \bibinfo {author} {\bibfnamefont {H.}~\bibnamefont {Takayasu}},\
  and\ \bibinfo {author} {\bibfnamefont {Y.}~\bibnamefont {Taguchi}},\
  }\bibfield  {title} {\bibinfo {title} {Non-{Gaussian} distribution in random
  transport dynamics},\ }\href@noop {} {\bibfield  {journal} {\bibinfo
  {journal} {International Journal of Modern Physics B}\ }\textbf {\bibinfo
  {volume} {8}},\ \bibinfo {pages} {3887} (\bibinfo {year} {1994})}\BibitemShut
  {NoStop}%
\bibitem [{\citenamefont {Takayasu}\ \emph {et~al.}(1996)\citenamefont
  {Takayasu}, \citenamefont {Kawakami}, \citenamefont {Taguchi},\ and\
  \citenamefont {Katsuyama}}]{takayasu1996fractal}%
  \BibitemOpen
  \bibfield  {author} {\bibinfo {author} {\bibfnamefont {H.}~\bibnamefont
  {Takayasu}}, \bibinfo {author} {\bibfnamefont {T.}~\bibnamefont {Kawakami}},
  \bibinfo {author} {\bibfnamefont {Y.}~\bibnamefont {Taguchi}},\ and\ \bibinfo
  {author} {\bibfnamefont {T.}~\bibnamefont {Katsuyama}},\ }\bibfield  {title}
  {\bibinfo {title} {Fractal limit distributions in random transports},\
  }\href@noop {} {\bibfield  {journal} {\bibinfo  {journal} {Fractals}\
  }\textbf {\bibinfo {volume} {4}},\ \bibinfo {pages} {257} (\bibinfo {year}
  {1996})}\BibitemShut {NoStop}%
\bibitem [{\citenamefont {Ispolatov}\ \emph {et~al.}(1998)\citenamefont
  {Ispolatov}, \citenamefont {Krapivsky},\ and\ \citenamefont
  {Redner}}]{ispolatov1998wealth}%
  \BibitemOpen
  \bibfield  {author} {\bibinfo {author} {\bibfnamefont {S.}~\bibnamefont
  {Ispolatov}}, \bibinfo {author} {\bibfnamefont {P.~L.}\ \bibnamefont
  {Krapivsky}},\ and\ \bibinfo {author} {\bibfnamefont {S.}~\bibnamefont
  {Redner}},\ }\bibfield  {title} {\bibinfo {title} {Wealth distributions in
  asset exchange models},\ }\href@noop {} {\bibfield  {journal} {\bibinfo
  {journal} {The European Physical Journal B-Condensed Matter and Complex
  Systems}\ }\textbf {\bibinfo {volume} {2}},\ \bibinfo {pages} {267} (\bibinfo
  {year} {1998})}\BibitemShut {NoStop}%
\bibitem [{\citenamefont {Dragulescu}\ and\ \citenamefont
  {Yakovenko}(2000)}]{dragulescu2000statistical}%
  \BibitemOpen
  \bibfield  {author} {\bibinfo {author} {\bibfnamefont {A.}~\bibnamefont
  {Dragulescu}}\ and\ \bibinfo {author} {\bibfnamefont {V.~M.}\ \bibnamefont
  {Yakovenko}},\ }\bibfield  {title} {\bibinfo {title} {Statistical mechanics
  of money},\ }\href@noop {} {\bibfield  {journal} {\bibinfo  {journal} {The
  European Physical Journal B-Condensed Matter and Complex Systems}\ }\textbf
  {\bibinfo {volume} {17}},\ \bibinfo {pages} {723} (\bibinfo {year}
  {2000})}\BibitemShut {NoStop}%
\bibitem [{\citenamefont {Heinsalu}\ and\ \citenamefont
  {Patriarca}(2014)}]{heinsalu2014kinetic}%
  \BibitemOpen
  \bibfield  {author} {\bibinfo {author} {\bibfnamefont {E.}~\bibnamefont
  {Heinsalu}}\ and\ \bibinfo {author} {\bibfnamefont {M.}~\bibnamefont
  {Patriarca}},\ }\bibfield  {title} {\bibinfo {title} {Kinetic models of
  immediate exchange},\ }\href@noop {} {\bibfield  {journal} {\bibinfo
  {journal} {The European Physical Journal B}\ }\textbf {\bibinfo {volume}
  {87}},\ \bibinfo {pages} {1} (\bibinfo {year} {2014})}\BibitemShut {NoStop}%
\bibitem [{\citenamefont {Van~Ginkel}\ \emph {et~al.}(2016)\citenamefont
  {Van~Ginkel}, \citenamefont {Redig},\ and\ \citenamefont
  {Sau}}]{van2016duality}%
  \BibitemOpen
  \bibfield  {author} {\bibinfo {author} {\bibfnamefont {B.}~\bibnamefont
  {Van~Ginkel}}, \bibinfo {author} {\bibfnamefont {F.}~\bibnamefont {Redig}},\
  and\ \bibinfo {author} {\bibfnamefont {F.}~\bibnamefont {Sau}},\ }\bibfield
  {title} {\bibinfo {title} {Duality and stationary distributions of the
  “immediate exchange model” and its generalizations},\ }\href@noop {}
  {\bibfield  {journal} {\bibinfo  {journal} {Journal of Statistical Physics}\
  }\textbf {\bibinfo {volume} {163}},\ \bibinfo {pages} {92} (\bibinfo {year}
  {2016})}\BibitemShut {NoStop}%
\bibitem [{\citenamefont {Barnsley}(2014)}]{barnsley2014fractals}%
  \BibitemOpen
  \bibfield  {author} {\bibinfo {author} {\bibfnamefont {M.~F.}\ \bibnamefont
  {Barnsley}},\ }\href@noop {} {\emph {\bibinfo {title} {Fractals
  Everywhere}}}\ (\bibinfo  {publisher} {Academic Press},\ \bibinfo {year}
  {2014})\BibitemShut {NoStop}%
\bibitem [{Note3()}]{Note3}%
  \BibitemOpen
  \bibinfo {note} {The implementation of the latter property explains why we
  choose $m/L$ rather than $t/L^2$ as second argument of
  $F_2(x,y)$.}\BibitemShut {Stop}%
\bibitem [{\citenamefont {Family}\ and\ \citenamefont
  {Vicsek}(1985)}]{FVscaling}%
  \BibitemOpen
  \bibfield  {author} {\bibinfo {author} {\bibfnamefont {F.}~\bibnamefont
  {Family}}\ and\ \bibinfo {author} {\bibfnamefont {T.}~\bibnamefont
  {Vicsek}},\ }\bibfield  {title} {\bibinfo {title} {Scaling of the active zone
  in the {E}den process on percolation networks and the ballistic deposition
  model},\ }\href@noop {} {\bibfield  {journal} {\bibinfo  {journal} {Journal
  of Physics A: Mathematical and General}\ }\textbf {\bibinfo {volume} {18}},\
  \bibinfo {pages} {L75} (\bibinfo {year} {1985})}\BibitemShut {NoStop}%
\bibitem [{\citenamefont {Sun}\ \emph {et~al.}(1989)\citenamefont {Sun},
  \citenamefont {Guo},\ and\ \citenamefont {Grant}}]{sun1989dynamics}%
  \BibitemOpen
  \bibfield  {author} {\bibinfo {author} {\bibfnamefont {T.}~\bibnamefont
  {Sun}}, \bibinfo {author} {\bibfnamefont {H.}~\bibnamefont {Guo}},\ and\
  \bibinfo {author} {\bibfnamefont {M.}~\bibnamefont {Grant}},\ }\bibfield
  {title} {\bibinfo {title} {Dynamics of driven interfaces with a conservation
  law},\ }\href@noop {} {\bibfield  {journal} {\bibinfo  {journal} {Phys. Rev.
  A}\ }\textbf {\bibinfo {volume} {40}},\ \bibinfo {pages} {6763} (\bibinfo
  {year} {1989})}\BibitemShut {NoStop}%
\bibitem [{\citenamefont {Krug}(1997)}]{krug1997origins}%
  \BibitemOpen
  \bibfield  {author} {\bibinfo {author} {\bibfnamefont {J.}~\bibnamefont
  {Krug}},\ }\bibfield  {title} {\bibinfo {title} {Origins of scale invariance
  in growth processes},\ }\href@noop {} {\bibfield  {journal} {\bibinfo
  {journal} {Advances in Physics}\ }\textbf {\bibinfo {volume} {46}},\ \bibinfo
  {pages} {139} (\bibinfo {year} {1997})}\BibitemShut {NoStop}%
\bibitem [{\citenamefont {Caballero}\ \emph {et~al.}(2018)\citenamefont
  {Caballero}, \citenamefont {Nardini}, \citenamefont {van Wijland},\ and\
  \citenamefont {Cates}}]{caballero2018strong}%
  \BibitemOpen
  \bibfield  {author} {\bibinfo {author} {\bibfnamefont {F.}~\bibnamefont
  {Caballero}}, \bibinfo {author} {\bibfnamefont {C.}~\bibnamefont {Nardini}},
  \bibinfo {author} {\bibfnamefont {F.}~\bibnamefont {van Wijland}},\ and\
  \bibinfo {author} {\bibfnamefont {M.~E.}\ \bibnamefont {Cates}},\ }\bibfield
  {title} {\bibinfo {title} {Strong coupling in conserved surface roughening: a
  new universality class?},\ }\href@noop {} {\bibfield  {journal} {\bibinfo
  {journal} {Phys. Rev. Lett.}\ }\textbf {\bibinfo {volume} {121}},\ \bibinfo
  {pages} {020601} (\bibinfo {year} {2018})}\BibitemShut {NoStop}%
\bibitem [{\citenamefont {Smith}\ \emph {et~al.}(2017)\citenamefont {Smith},
  \citenamefont {Meerson},\ and\ \citenamefont {Sasorov}}]{Smith_etal-17}%
  \BibitemOpen
  \bibfield  {author} {\bibinfo {author} {\bibfnamefont {N.~R.}\ \bibnamefont
  {Smith}}, \bibinfo {author} {\bibfnamefont {B.}~\bibnamefont {Meerson}},\
  and\ \bibinfo {author} {\bibfnamefont {P.~V.}\ \bibnamefont {Sasorov}},\
  }\bibfield  {title} {\bibinfo {title} {Local average height distribution of
  fluctuating interfaces},\ }\href@noop {} {\bibfield  {journal} {\bibinfo
  {journal} {Phys. Rev. E}\ }\textbf {\bibinfo {volume} {95}},\ \bibinfo
  {pages} {012134} (\bibinfo {year} {2017})}\BibitemShut {NoStop}%
\bibitem [{\citenamefont {Majumdar}\ \emph {et~al.}(1998)\citenamefont
  {Majumdar}, \citenamefont {Krishnamurthy},\ and\ \citenamefont
  {Barma}}]{Majumdar_etal-98}%
  \BibitemOpen
  \bibfield  {author} {\bibinfo {author} {\bibfnamefont {S.~N.}\ \bibnamefont
  {Majumdar}}, \bibinfo {author} {\bibfnamefont {S.}~\bibnamefont
  {Krishnamurthy}},\ and\ \bibinfo {author} {\bibfnamefont {M.}~\bibnamefont
  {Barma}},\ }\bibfield  {title} {\bibinfo {title} {Nonequilibrium phase
  transitions in models of aggregation, adsorption, and dissociation},\ }\href
  {https://doi.org/10.1103/PhysRevLett.81.3691} {\bibfield  {journal} {\bibinfo
   {journal} {Phys. Rev. Lett.}\ }\textbf {\bibinfo {volume} {81}},\ \bibinfo
  {pages} {3691} (\bibinfo {year} {1998})}\BibitemShut {NoStop}%
\bibitem [{\citenamefont {Rajesh}\ and\ \citenamefont
  {Majumdar}(2001)}]{Rajesh_Majumdar_2001}%
  \BibitemOpen
  \bibfield  {author} {\bibinfo {author} {\bibfnamefont {R.}~\bibnamefont
  {Rajesh}}\ and\ \bibinfo {author} {\bibfnamefont {S.~N.}\ \bibnamefont
  {Majumdar}},\ }\bibfield  {title} {\bibinfo {title} {Exact phase diagram of a
  model with aggregation and chipping},\ }\href@noop {} {\bibfield  {journal}
  {\bibinfo  {journal} {Physical Review E}\ }\textbf {\bibinfo {volume} {63}},\
  \bibinfo {pages} {036114} (\bibinfo {year} {2001})}\BibitemShut {NoStop}%
\bibitem [{\citenamefont {Falcon}\ and\ \citenamefont
  {Mordant}(2022)}]{Falcon-Mordant-22}%
  \BibitemOpen
  \bibfield  {author} {\bibinfo {author} {\bibfnamefont {E.}~\bibnamefont
  {Falcon}}\ and\ \bibinfo {author} {\bibfnamefont {N.}~\bibnamefont
  {Mordant}},\ }\bibfield  {title} {\bibinfo {title} {Experiments in surface
  gravity–capillary wave turbulence},\ }\href
  {https://doi.org/10.1146/annurev-fluid-021021-102043} {\bibfield  {journal}
  {\bibinfo  {journal} {Annual Review of Fluid Mechanics}\ }\textbf {\bibinfo
  {volume} {54}},\ \bibinfo {pages} {1} (\bibinfo {year} {2022})},\ \Eprint
  {https://arxiv.org/abs/https://doi.org/10.1146/annurev-fluid-021021-102043}
  {https://doi.org/10.1146/annurev-fluid-021021-102043} \BibitemShut {NoStop}%
\bibitem [{\citenamefont {Peres}\ \emph {et~al.}(2000)\citenamefont {Peres},
  \citenamefont {Schlag},\ and\ \citenamefont {Solomyak}}]{peres2000sixty}%
  \BibitemOpen
  \bibfield  {author} {\bibinfo {author} {\bibfnamefont {Y.}~\bibnamefont
  {Peres}}, \bibinfo {author} {\bibfnamefont {W.}~\bibnamefont {Schlag}},\ and\
  \bibinfo {author} {\bibfnamefont {B.}~\bibnamefont {Solomyak}},\ }\bibfield
  {title} {\bibinfo {title} {Sixty years of {B}ernoulli convolutions},\ }in\
  \href@noop {} {\emph {\bibinfo {booktitle} {Fractal geometry and stochastics
  II}}}\ (\bibinfo  {publisher} {Springer},\ \bibinfo {year} {2000})\ pp.\
  \bibinfo {pages} {39--65}\BibitemShut {NoStop}%
\bibitem [{\citenamefont {Fraser}\ and\ \citenamefont
  {Kapral}(1992)}]{fraser1992periodic}%
  \BibitemOpen
  \bibfield  {author} {\bibinfo {author} {\bibfnamefont {S.~J.}\ \bibnamefont
  {Fraser}}\ and\ \bibinfo {author} {\bibfnamefont {R.}~\bibnamefont
  {Kapral}},\ }\bibfield  {title} {\bibinfo {title} {Periodic
  dichotomous-noise-induced transitions and stochastic coherence},\ }\href@noop
  {} {\bibfield  {journal} {\bibinfo  {journal} {Physical Review A}\ }\textbf
  {\bibinfo {volume} {45}},\ \bibinfo {pages} {3412} (\bibinfo {year}
  {1992})}\BibitemShut {NoStop}%
\bibitem [{\citenamefont {Kapral}\ and\ \citenamefont
  {Fraser}(1993)}]{kapral1993dynamics}%
  \BibitemOpen
  \bibfield  {author} {\bibinfo {author} {\bibfnamefont {R.}~\bibnamefont
  {Kapral}}\ and\ \bibinfo {author} {\bibfnamefont {S.~J.}\ \bibnamefont
  {Fraser}},\ }\bibfield  {title} {\bibinfo {title} {Dynamics of oscillators
  with periodic dichotomous noise},\ }\href@noop {} {\bibfield  {journal}
  {\bibinfo  {journal} {J. Stat. Phys.}\ }\textbf {\bibinfo {volume} {70}},\
  \bibinfo {pages} {61} (\bibinfo {year} {1993})}\BibitemShut {NoStop}%
\bibitem [{\citenamefont {Peres}\ \emph {et~al.}(2006)\citenamefont {Peres},
  \citenamefont {Simon},\ and\ \citenamefont {Solomyak}}]{peres2006absolute}%
  \BibitemOpen
  \bibfield  {author} {\bibinfo {author} {\bibfnamefont {Y.}~\bibnamefont
  {Peres}}, \bibinfo {author} {\bibfnamefont {K.}~\bibnamefont {Simon}},\ and\
  \bibinfo {author} {\bibfnamefont {B.}~\bibnamefont {Solomyak}},\ }\bibfield
  {title} {\bibinfo {title} {Absolute continuity for random iterated function
  systems with overlaps},\ }\href@noop {} {\bibfield  {journal} {\bibinfo
  {journal} {Journal of the London Mathematical Society}\ }\textbf {\bibinfo
  {volume} {74}},\ \bibinfo {pages} {739} (\bibinfo {year} {2006})}\BibitemShut
  {NoStop}%
\bibitem [{\citenamefont {Barral}\ and\ \citenamefont
  {Feng}(2021)}]{barral2021multifractal}%
  \BibitemOpen
  \bibfield  {author} {\bibinfo {author} {\bibfnamefont {J.}~\bibnamefont
  {Barral}}\ and\ \bibinfo {author} {\bibfnamefont {D.-J.}\ \bibnamefont
  {Feng}},\ }\bibfield  {title} {\bibinfo {title} {On multifractal formalism
  for self-similar measures with overlaps},\ }\href@noop {} {\bibfield
  {journal} {\bibinfo  {journal} {Mathematische Zeitschrift}\ }\textbf
  {\bibinfo {volume} {298}},\ \bibinfo {pages} {359} (\bibinfo {year}
  {2021})}\BibitemShut {NoStop}%
\bibitem [{\citenamefont {Pikovsky}\ and\ \citenamefont
  {Tsimring}(2023)}]{Pikovsky-Tsimring-23}%
  \BibitemOpen
  \bibfield  {author} {\bibinfo {author} {\bibfnamefont {A.}~\bibnamefont
  {Pikovsky}}\ and\ \bibinfo {author} {\bibfnamefont {L.~S.}\ \bibnamefont
  {Tsimring}},\ }\bibfield  {title} {\bibinfo {title} {Statistical theory of
  asymmetric damage segregation in clonal cell populations},\ }\href@noop {}
  {\bibfield  {journal} {\bibinfo  {journal} {Mathematical Biosciences}\
  }\textbf {\bibinfo {volume} {358}},\ \bibinfo {pages} {108980} (\bibinfo
  {year} {2023})}\BibitemShut {NoStop}%
\end{thebibliography}%

\end{document}